\documentclass[journal]{IEEEtran}
\usepackage{}

\ifCLASSINFOpdf
\else
\fi
\usepackage{stfloats}
\usepackage{amsfonts}
\usepackage{amssymb}
\usepackage{amsmath}
\usepackage{mathrsfs}
\usepackage{cite}
\usepackage{subfigure}
\usepackage{graphicx}
\usepackage{algorithm}
\usepackage{enumitem}

\usepackage{empheq}

\usepackage{bbm}
\usepackage{leftidx}
\usepackage{extarrows}
\usepackage{colortbl}
\usepackage{CJK}

\usepackage{enumitem}

\usepackage{algorithm}
\usepackage{algorithmic} 

\usepackage{empheq}

\newtheorem{thm}{Theorem}

\newtheorem{cor}{Corollary}
\newtheorem{lem}{Lemma}
\newtheorem{rmk}{Remark}
\newtheorem{de}{Definition}
\newtheorem{exa}{Example}
\newtheorem{pro}{Proposition}

\newcommand*{\QEDA}{\hfill\ensuremath{\blacklozenge}}  

\usepackage{multicol} 
\newlength{\halfpagewidth}
\divide\halfpagewidth by 2
\newcommand{\leftsep}{%
\noindent\raisebox{4mm}[0ex][0ex]{%
\makebox[\halfpagewidth]{\hrulefill}\hbox{\vrule height 3pt}}%
\vspace*{-2mm}%
}
\newcommand{\rightsep}{%
\noindent\hspace*{\halfpagewidth}%
\rlap{\raisebox{-3pt}[0ex][0ex]{\hbox{\vrule height 3pt}}}%
\makebox[\halfpagewidth]{\hrulefill}%
}

\begin{document}
%
\title{A Two-Functional-Network Framework of Opinion Dynamics}
%
%
%

\author{Wentao~Zhang,
Zhiqiang~Zuo,
 and Yijing~Wang
\thanks{This work was supported by the National Natural Science Foundation of China No. 61933014, No. 61773281, No. 61673292.}
\thanks{
 The authors are with the Tianjin Key
Laboratory of Process Measurement and Control, School of Electrical and Information Engineering, Tianjin University, Tianjin, 300072, P. R. China.
(e-mail: \{wtzhangee, zqzuo, yjwang\}@tju.edu.cn).}}
\maketitle

\begin{abstract}
A common trait involving the opinion dynamics in social networks is an anchor on interacting network to characterize the opinion formation process among participating social actors, such as information flow, cooperative and antagonistic influence, etc. Nevertheless, interacting networks are generally public for social groups, as well as other individuals who may be interested in. This blocks a more precise interpretation of the opinion formation process since social actors always have complex feeling, motivation and behavior, even beliefs that are personally private. In this paper, we formulate a general configuration on describing how individual's opinion evolves in a distinct fashion. It consists of two functional networks: interacting network and appraisal network. Interacting network inherits the operational properties as DeGroot iterative opinion pooling scheme while appraisal network, forming a belief system, quantifies certain cognitive orientation to interested individuals' beliefs, over which the adhered attitudes may have the potential to be antagonistic. We explicitly show that cooperative appraisal network always leads to consensus in opinions. Antagonistic appraisal network, however, causes opinion cluster. It is verified that antagonistic appraisal network affords to guarantee consensus by imposing some extra restrictions. They hence bridge a gap between the consensus and the clusters in opinion dynamics. We further attain a gauge on the appraisal network by means of the random convex optimization approach. Moreover, we extend our results to the case of mutually interdependent issues.
\end{abstract}


%
\IEEEpeerreviewmaketitle

\section{Introduction}\label{sec1}
The study of opinion dynamics in social networks has a long history. Compared with many natural or man-made systems (networks), social actors (agents) in social networks rarely display a common interest. In contrast, they usually aggregate into several small groups where the agents in the same group achieve a unanimous behavior while the opinions of the whole network comprise several clusters. Opinion dynamics in social networks are universal topics and have captured massive interests from different disciplines, such as sociology, social anthropology, economics, ideological political science, physics, biology and control theory, even in the field of military \cite{friedkin2011social,FB-LNS}.

A simple yet instructive mathematical model is fundamental for the study of opinion dynamics in social networks. As a backbone for opinion dynamics, DeGroot's iterative opinion pooling configuration shows that social actors can share a common viewpoint if a convex combination mechanism is performed (cf. \cite{Degroot1974reaching}). In many practical situations, agents often interact with those who are like-minded, and agree on more deviant viewpoints with discretion. For Hegselmann-Krause model, each agent only communicates with those whose opinions are constrained into its confidence interval \cite{hegselmann2002opinion,blondel2009krause}. Actually, this model is implicitly based on the principle of biased assimilation or homophily. It is always the case that some individuals in social networks have their own prejudices, no matter what kind of opinion formation mechanism is applied. An attempt towards this direction is the Friedkin-Johnsen (FJ) model where some of the individuals (stubbornness) are affected by external signal \cite{friedkin1999social}. Unlike Hegselmann-Krause model, FJ model achieves opinion clusters even in the linear opinion formation process. Conventionally, FJ model mainly focuses on the issue free opinion dynamics or the scalar opinion dynamics. Refs. \cite{friedkin2015problem,fortunato2005vector} extended the FJ model to the vector-based opinion dynamics where the opinions tightly relate to several issues, see also \cite{friedkin2017truth}. As hinted by belief systems (cf. \cite{converse2006nature}), topic-specific
opinion dynamics are usually entwined whenever agent's opinion is involved into several interdependent issues. Parsegov \emph{et al}. \cite{parsegov2017novel} introduced a row stochastic matrix (MiDS matrix) which clarifies what the attitude of agent towards the issue sequence is. Similarly, the property of belief system dynamics subject to logical constraints was elaborated in \cite{friedkin2016network}, and a case study was provided to show how the fluctuations of population's attitudes evolve. The gossip-based version of the FJ model was proposed in \cite{ravazzi2015ergodic}. More recently, an approach building on the DeGroot model and the FJ model was given to investigate the evolutions with respect to the self-appraisal, interpersonal influences and social power over issue sequence for the star, the irreducible and the reducible communication topologies \cite{jia2015opinion}.

Nowadays, people have realized  that postulate of ``cognitive algebra" on heterogeneous information (cf. \cite{anderson1981foundations}) for opinion formation process among individuals may not attribute all complex behaviors in social networks due to the negative interpersonal influences that often emerge in many community of dissensus (cf. \cite{petty2012communication,flache2011small}). Typical instances include but are not limited to multiple-party political systems, biological systems, international alliances, bimodal coalitions, rival business cartels, duopolistic markets, as well as boomerang effect in dyad systems. Based on multi-agent system theory, a lot of research interests have been devoted to the opinion polarization and the stabilization problems. By signed graph theory, Altafini \cite{altafini2013consensus} proved that structurally balanced graph is the necessary and sufficient condition for bipartite consensus. Proskurnikov \emph{et al}. \cite{proskurnikov2016opinion} further reported the necessary and sufficient criterion to stabilize Altafini's model. For general communication graph, \cite{meng2016interval} discussed the interval bipartite consensus problem. The second-order/high-order multi-agent systems in the presence of antagonistic information were also discussed in \cite{zhang2020nonlinear,valcher2014consensus}, as well as finite-time consensus \cite{lu2019finite}. More related work about the signed graph theory refers to \cite{proskurnikov2018tutorial,shi2019dynamics}. Cooperative control with antagonistic reciprocity was discussed in \cite{zhang2017cooperative,zhang2017quasi,zhang2019double} using the node-based mechanism.

Unlike complex systems or multi-agent systems, agents (individuals) in social networks usually display diversity behaviors, such as attitude, feeling and belief, etc. In business negotiation or alliance, for instance, each member tends to collect more ``sides information" with respect to the remaining members before organizing a meeting, by which they aim at achieving the profit maximization. They possess introspective ability to access what and how they should do. Notice that the above research is based on a common hypothesis that merely the public network is used to quantify opinion formation process. However, a downside of public networks\footnote{As hinted by \cite{friedkin2011social} and \cite{parsegov2017novel}, we treat the interaction topologies (communication graphs) in the aforementioned literature by the public networks.} in the aforementioned literature to describe the antagonistic interactions among participating individuals is that individuals in social networks can completely get access to the opinion, the attitude and the belief of the other individuals towards the interested individuals. But this is not always the case since the individuals in social networks feature complex behaviors, even complex thoughts \cite{wang2018opinion}, naturally leading to the complex opinion dynamics that the natural or the man-made complex networks cannot show. Moreover, as discussed in \cite{friedkin2006structural,parsegov2017novel}, public networks generally characterize the social influence structure, and henceforth they are assumed to be thoroughly known. For this reason, potential hostile information will become transparent whenever only the public network is utilized to model the opinion formation process. Obviously, this is not always true for social networks where it is rather arduous to have access to the individual's opinion in prior.

In this paper, we propose a new framework with functional networks, that is, the appraisal network and the interacting network, to describe the opinion formation process. The opinion evolution in social network is firstly governed by an appraisal network characterizing how each individual assigns its attitude or influence towards other individuals. Afterwards, each individual updates its opinion through an interacting network as the conventional DeGroot model (we call it the interacting network or the public network). To the best of the authors' knowledge, there are no results available to model the evolution of opinion dynamics using two functionally independent networks. More importantly, we will show that the proposed formulation indeed provides some intriguing phenomena that cannot be preserved by the existing setups. We summarize the main features of this paper as follows:
\begin{description}
  \item[(a)] Cooperative appraisal networks lead to consensus in opinion whereas the final aggregated value may not be contained in a convex hull spanned by the initial opinions of the participating individuals. This potentially implies that the way of the decision among individuals in social networks formulated in this paper is not necessarily constrained into the convex hull as the usual models. In fact, non-convex interactions in social networks are common, and it is natural to model the evolution of the opinions by taking this factor into account \cite{wang2018opinion};
  \item[(b)] Antagonistic appraisal networks result in clusters in opinions. In particular, we show that consensus in opinion dynamics appears provided that the considered antagonistic appraisal networks enjoy certain requirements. It is illustrated that most of the existing results merely guarantee the clusters in opinions subject to hostile interactions (cf. \cite{proskurnikov2018tutorial,shi2019dynamics}), or stability of the agents \cite{proskurnikov2016opinion};


  \item[(c)] Random convex optimization (cf. \cite{calafiore2006scenario,calafiore2010random}) is formulated to provide a feasible estimation on the appraisal network, with the purpose of finite-horizon sense. Also, a bound on the number of ``required observations", which enables us to get rid of a-prior specified level of probabilistic violation, is explicitly given. Therefore we can make a justification of the postulate on self-preservation of the appraisal network for each individual, as opposed to the hypothesis on interacting network in the literature;
  \item[(d)] The proposed setup could be further extended to the multiple mutually entangled issues that are quantified by an issue-dependence matrix, upon which we deduce the criteria associated with the convergence (resp. stability) of the agents. More interestingly, we point out that the introduction of the issue-dependence matrix enables us to steer the leader's opinion, which has been verified to be fixed all the time in the context of multi-agent systems community.
\end{description}

%
%


The layout of this paper is outlined as follows: Section ${\rm \ref{20191 sec2}}$ describes some basic preliminaries and the problem
formulation as well as the dynamic model. Section ${\rm \ref{20191 sec3}}$ presents the results in the context of the cooperative and the antagonistic appraisal networks. Section ${\rm \ref{20191 sec4}}$ reports the results for the interdependent issues. Section ${\rm \ref{20191 sec5}}$ gives numerical examples to support the derived theoretical results as well as some discussions. Finally, a conclusion is drawn in Section ${\rm \ref{20191 sec6}}$.


 \section{Preliminaries and Problem Formulation}\label{20191 sec2}
\subsection{Basic Notations}
The real set, the nonnegative real set and the nonnegative integer set are denoted by $\mathbb{R}$, $\mathbb{R}_{+}$ and $\mathbb{Z}$. Symbol $``\prime"$ denotes the transpose regarding a vector or a matrix. Symbol $|\cdot |$ represents the modulus or the cardinality, and $| \cdot |_{2}$ the $2$-norm. $|Q|$ stands for the spectral norm for matrix $Q$. $\lambda(Q)$ and $Q^{-1}$ denote the eigenvalue and the inverse of an invertible matrix $Q$. $Q\succ0$ indicates that matrix $Q$ is symmetric, and positive definite. We denote $\{1,...,N\}$ and $(1,...,1)^{\prime} $ by $\mathbb{I}_{N}$ and $\textbf{1}_{N}$, where $N$ is the number of agents in social networks. $I$ and $\mathcal {O}$ are, respectively, the identity matrix and the zero matrix. The diagonal matrix is represented by ${\rm diag}(\cdot)$. The sign function is abbreviated by ${\rm sgn}(\cdot)$. For any complex number $\lambda$, $\lambda\triangleq{\rm Re}(\lambda)+\mathbbm{i}{\rm Im}(\lambda)=|\lambda|(\cos(\arg(\lambda))+\mathbbm{i}\sin(\arg(\lambda))) $ where $\mathbbm{i}^{2}=-1$ and $\arg(\lambda)$ represents the value of argument principle.

In addition, we denote $(\Omega, \mathcal{F}, \mathcal{P})$ the probability space, upon which $\Omega$ represents the sample space, $\mathcal{F}$ the Borel $\sigma$-algebra, and $\mathcal{P}$ the probability measure.

An interacting graph $\mathcal{G}$ is commonly represented by a triple $\{\mathbb{V},\mathbb{E},
\mathbb{A}\}$ where $\mathbb{V}$ is the node set, $\mathbb{E}$ is the edge set and $\mathbb{A}=(a_{ij})_{N\times N}$ is the adjacent matrix with $a_{ij}>0$ provided that $(j,i) \in \mathbb{E}$ and $a_{ij}=0$ otherwise. No self-loop is allowed throughout the  paper, i. e.,  $a_{ii}=0$. The associated Laplacian matrix $\mathcal{L}=(l_{ij})_{N\times N}$ is given by $l_{ii}=\sum^{N}_{j=1,j\neq i}a_{ij}$ and $l_{ij,j\neq i}=-a_{ij,j\neq i}$\footnote{In the context of the opinion dynamics, $p_{ij}$ represents the attitude of agent $i$ towards agent $j$, while $l_{ij}$ in the framework of the multi-agent systems means that agent $j$ is a neighbor of agent $i$. Therefore, if no confusion arises, we treat both $p_{ij}$ and $l_{ij}$ as the influence of agent $j$ imposed on agent $i$ with the purpose of the notion consistency.}.
A digraph is strongly connected if for any two distinct nodes, they are connected by a path. A root of $\mathcal{G}$ is a special node, from which we can arrive at any other nodes. A graph has a directed spanning tree if and only if it contains at least a root.

\subsection{Dynamic Model for Social Networks}
Consider a group of interacting individuals (agents or social actors), whose opinions evolve according to
\begin{subequations}\label{20191eq1}
\begin{empheq}[left=\empheqlbrace]{align}
\xi_{i}(k+1)=&~\xi_{i}(k)+\varrho_{i}\sum_{j\in\mathcal{N}_{i}}  a_{ij}(z_{j}(k)-z_{i}(k)) \label{20191eq1a}\\
z_{i}(k)=&~\sum^{N}_{j=1}\delta_{ij}\xi_{j}(k),~i\in\mathbb{I}_{N},  ~k\in \mathbb{Z}  \label{20191eq1b}
\end{empheq}
\end{subequations}
where $\xi_{i}(k)\in \mathbb{R}$ stands for the opinion of the $i$th agent at instant $k$, $\mathcal{N}_{i}$ is the neighboring set of agent $i$. Similar to the FJ model, we call parameter $\varrho_{i}$ ($\varrho_{i}\neq 0$) the susceptibility factor with respect to the $i$th social actor. $z_{i}(k)$ in $(\ref{20191eq1b})$ represents the appraisal or the self-appraisal ($z_{i}(k)=\delta_{i}\xi_{i}(k)$) of agent $i$. The constant $\delta_{ij}$ specifies the weighted influence assigned by individual $i$ towards individual $j$, and satisfies $0<\sum^{N}_{j=1}|\delta_{ij}|\leq 1$. The compact form of $(\ref{20191eq1})$ is expressed by
\begin{equation}\label{20191eq2}
\begin{aligned}
\xi(k+1)=&~(I-\Lambda\mathcal {L}\mathcal {D})\xi(k),~k\in \mathbb{Z}
\end{aligned}
\end{equation}
where $\xi(k)=(\xi_{1}(k),...,\xi_{N}(k))^{\prime}$, $\Lambda={\rm diag}(\varrho_{1},...,\varrho_{N})$ and $\mathcal {D}=(\delta_{ij})_{N\times N}$.

It is easy to see that system $(\ref{20191eq2})$ involves two networks, and we call them the interacting or the public network (quantified by matrix $\mathcal {L}$) and the appraisal network (quantified by matrix $\mathcal {D}$), respectively. We emphasize that the susceptibility factor $\varrho_{i}$ is crucial for the convergence of system $(\ref{20191eq2})$ since the appraisal network and the interacting network are entwined, leading to a significant difference in contrast with the DeGroot model. It is noted that a system might collapse due to the hostility. And as will be shown later, a system subject to antagonistic information fails to converge, regardless of the connection property of the underlying interacting graph\footnote{For Ref. \cite{altafini2013consensus}, the proposed consensus algorithm always assures the convergence of the reciprocal agents as long as the communication topology attains a directed spanning tree, that is, bipartite consensus (interval bipartite consensus) for structurally balanced graph (cf.\cite{altafini2013consensus,meng2016interval}) and stability for graph that contains the in-isolated structurally balanced subgraphs, or is structurally unbalanced (cf. \cite{proskurnikov2016opinion}).}.

\begin{rmk}
One can easily see that system $(\ref{20191eq2})$ boils down to the classical DeGroot model (cf. \cite{Degroot1974reaching}) or the multi-agent systems (cf. \cite{olfati2004consensus}) when we fix $\mathcal {D}$ to be an identity matrix and $\varrho_{i}$ some positive constant $\varrho$ fulfilling $0<\varrho< \max_{i\in \mathbb{I}_{N}} \{l_{ii}\}$ (in such a scenario we always treat $\varrho$ as the step size). It should be pointed out that the two networks described above have nothing to do with the multilayer networks in the context of complex networks \cite{de2013mathematical} where all layer networks functionally inherent with the interacting network $\mathcal {L}$. \QEDA
\end{rmk}

We are now in a position to give some interpretations about the reason why we introduce the appraisal network: $\textbf{(1)}$ Unlike the natural and the man-made complex networks where the interacting agents are creatures and smart machines, the individuals in social networks are people. In one word, the subject in social networks possesses emotion, belief as well as attitude towards a specific object. Therefore, it is a common sense that people try to search and collect as much information as possible before they make a decision, such as organizing a conference, exchanging ideas with colleagues, etc. That is to say, people in social networks always
evaluate and self-reflect their opinions, beliefs, and behaviors before interacting with others; $\textbf{(2)}$ As mentioned before, the hostile interaction is a key element in the study of opinion dynamics in social networks. It is notable that the interacting networks are generally assumed to be completely known (cf. \cite{friedkin2006structural,parsegov2017novel}). Therefore, the private opinions of the participating agents may be leaked if merely the interacting networks are applied to model the opinion evolution in social networks, especially, involving antagonistic interactions.

Before moving on, we give some useful definitions regarding system $(\ref{20191eq2})$.
\begin{de}\label{20191de1}
System $(\ref{20191eq2})$ achieves:
\begin{itemize}
\item[({\rm \lowercase \expandafter {\romannumeral 1}})] the \emph{consensus in opinions}, if
\begin{equation*}\label{20191eq3}
\begin{aligned}
\lim _{k\rightarrow\infty}\xi_{i}(k)=\varphi,~~i\in\mathbb{I}_{N}
\end{aligned}
\end{equation*}
where $\varphi\in \mathbb{R}$ is a constant.
\item[({\rm \lowercase \expandafter {\romannumeral 2}})] the \emph{convergence in opinions}, if
\begin{equation*}\label{20191eq4}
\begin{aligned}
\lim _{k\rightarrow\infty}\xi_{i}(k)=\varphi_{i},~~i\in\mathbb{I}_{N}
\end{aligned}
\end{equation*}
where $\varphi_{i}\in \mathbb{R}$ is a constant.
\item[({\rm \lowercase \expandafter {\romannumeral 3}})] the \emph{stability in opinions}, if
\begin{equation*}\label{20191eq65}
\begin{aligned}
\lim _{k\rightarrow\infty}\xi_{i}(k)=0,~~i\in\mathbb{I}_{N}
\end{aligned}
\end{equation*}
\end{itemize}
\end{de}

The mechanism behind Definition $\ref{20191de1}$ is that a cooperative appraisal network gives rise to the opinion aggregation while an antagonistic appraisal network leads to the clusters in opinion in general. Moreover, the consensus of the reciprocal agents is a special type of the convergence where all interacting individuals share a common viewpoint eventually.

\begin{figure}
\centering
\includegraphics[width=3.45in,height=1.1in]{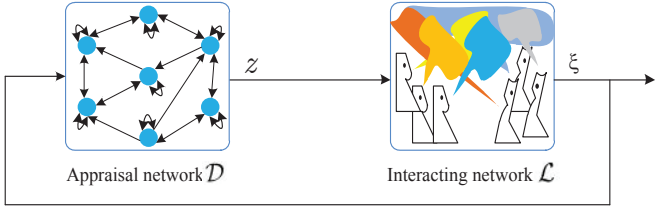}
\caption{A holistic paradigm of the opinion evolution for system $(\ref{20191eq2})$ where the appraisal network (characterized by $\mathcal {D}$) is cooperative, i.e., $\delta_{ij}\geq 0$.}
\label{20191_fig1_1}
\end{figure}

\section{Main Results}\label{20191 sec3}

In the following discussions, we will show that the opinion formation process with setup $(\ref{20191eq1})$ is more  general comparing with the DeGroot model. More specific, system $(\ref{20191eq1})$ permits non-convex interactions of the individuals, and it provides an intriguing viewpoint on how the individual's introspective process influences the opinion evolution of social networks.

\subsection{Cooperative Appraisal Network}\label{20191 sec41}
In this subsection, we are concerned with the situation where the appraisal network is cooperative, i.e., each influence weight $\delta_{ij}$ is nonnegative.  A simple illustration for this case is drawn in Fig. $\ref{20191_fig1_1}$. Hence, $\sum^{N}_{j=1}\delta_{ij}=1$. In such a case, $\mathcal {D}=(\delta_{ij})_{N\times N}$ is a nonnegative stochastic matrix. Before proceeding on, we introduce a simple proposition which bridges the gap between the stochastic matrix and the Laplacian matrix in connection with the communication topology in a unified manner.
\begin{pro}(\hspace{-0.001cm}\cite{olfati2004consensus})\label{20191pro1}
Given any nonnegative stochastic matrix $\mathcal {D}=(\delta_{ij})_{N\times N}$, there always exists a Laplacian matrix $L=(l_{ij})_{N\times N}$ such that
\begin{equation}\label{20191eq37}
\begin{aligned}
\mathcal {D}=I-\epsilon L
\end{aligned}
\end{equation}
where $\epsilon>0$ is the step size.
\end{pro}


It should be emphasized that the Laplacian matrix $L$ depicted in $(\ref{20191eq37})$ generally has little connection with the interacting network $\mathcal{L}$ reported in $(\ref{20191eq2})$. In the sequel, we give the main results on cooperative appraisal network.

\begin{thm}\label{20191th1}
Suppose that the appraisal network is cooperative, i.e., $\delta_{ij} \geq 0$. Then \emph{consensus} in system $(\ref{20191eq2})$ is achieved if and only if matrix $I-\Lambda\mathcal {L}\mathcal {D}$ has a simple $1$ eigenvalue and the remaining eigenvalues are preserved in the unit disk.
\end{thm}



\begin{IEEEproof}
Suppose that the communication graph associated with $\mathcal {L}$ has a directed spanning tree. By Sylvester rank inequality, one has
\begin{equation*}\label{20191eq5}
\begin{aligned}
{\rm rank}(\mathcal {L})+{\rm rank}(\mathcal {D})-  N \leq & ~{\rm rank}(\mathcal {L}\mathcal {D})\\
\leq& ~\min \bigg\{{\rm rank}(\mathcal {L}),{\rm rank}(\mathcal {D})\bigg\}
\end{aligned}
\end{equation*}
In fact, the zero eigenvalue is simple if the graph induced by $\mathcal {L}$ has a directed spanning tree. Then, zero is an eigenvalue of matrix $\Lambda\mathcal {L}\mathcal {D}$. Additionally, vector one is the eigenvector of the zero eigenvalue. Hence, system $(\ref{20191eq2})$ can achieve the consensus.

We proceed with the condition for consensus. Actually, matrix $I-\Lambda\mathcal {L}\mathcal {D}$ has a simple $1$ eigenvalue and the remaining eigenvalues are preserved in the unit disk if and only if $\Lambda\mathcal {L}\mathcal {D}$ contains a simple zero eigenvalue and the remaining eigenvalues share positive real parts (here we can redefine the coupling matrix by $\Lambda=\epsilon_1\Lambda^{\dag}$ where $\epsilon_1$ is a small step size, by doing so we could access to the continuous version of $(\ref{20191eq2})$ by $\dot{\xi}=-\Lambda^{\dag}\mathcal {L}\mathcal {D}\xi$). It further assumes that the left and the right eigenvectors with respect to the zero eigenvalue are, respectively, $\varsigma\in \mathbb{R}^{N}$ and $\iota \in \mathbb{R}^{N}$ with the constraint $\varsigma^{\prime}\iota=1$. Note that $\iota=\textbf{1}_{N}$ for such a scenario since $\mathcal {D}\iota=\iota$. We define the disagreement error by $\theta(k)=(I-\iota\varsigma^{\prime})\xi(k)$, then one has
\begin{equation}\label{20191eq11}
\begin{aligned}
\theta(k+1)=(I-\Lambda\mathcal {L}\mathcal {D})\theta(k)
\end{aligned}
\end{equation}
We can see that the eigenvalues of $I-\Lambda\mathcal {L}\mathcal {D}$ are entirely constrained in the unit disk over the space $\mathbb{R}^{N}\backslash \{ \varphi\}$ where $\varphi=\iota\varsigma^{\prime}\xi(0)=\alpha \iota$. Thus the error system $(\ref{20191eq11})$ is exponentially stable, which ensures the consensus of system $(\ref{20191eq2})$. This ends the proof by Definition $\ref{20191de1}$.
\end{IEEEproof}

\begin{rmk}
Different from the DeGroot model\cite{Degroot1974reaching} and multi-agent systems\cite{olfati2004consensus}, the final shared common opinion of the interacting individuals governed by $(\ref{20191eq2})$ may not be restricted in a convex hull spanned by the initial opinions of the individuals. This is because that the appraisal network specifying the interaction rule in this paper is no long to be convex, to a large extent. We emphasize that non-convex interactions among participating individuals in social networks are rather common, which have been clearly pointed out by Wang \emph{et al}. in \cite{wang2018opinion}. This intriguing issue will be verified via an example later.\QEDA
\end{rmk}

Next, we consider a special case where the appraisal network is the same as the interacting network, i.e., $L=\mathcal {L}$.
In order to achieve the consensus, it suffices to fix $\varrho_{i}=\varrho$ for all $i$. One hence arrives at
\begin{equation}\label{20191eq6}
\begin{aligned}
\xi(k+1)=&~(I-\varrho\mathcal {L}+\epsilon\varrho\mathcal {L}^{2})\xi(k)
\end{aligned}
\end{equation}
where $\epsilon>0$ is a constant.

\begin{cor}\label{20191cor1}
Suppose that the graph induced by $\mathcal {L}$ has a directed spanning tree. Then system $(\ref{20191eq6})$ achieves the \emph{consensus in opinions} if and only if
\begin{equation}\label{20191eq10}
\begin{aligned}
0< \varrho< \min _{{\rm Re}(\lambda^{\star}_{i})>0}\bigg\{\frac{2{\rm Re}(\lambda^{\star}_{i})}{{\rm Re}^{2}(\lambda^{\star}_{i})+{\rm Im}^{2}(\lambda^{\star}_{i})}\bigg\}
\end{aligned}
\end{equation}
where $\lambda^{\star}_{i}$ stands for the $i$th eigenvalue of $\mathcal {L}-\epsilon\mathcal {L}^{2}$, and constant $\epsilon$ satisfies
\begin{equation}\label{ch7_20191eq10_1}
\left\{\begin{aligned}
&\epsilon\in \mathbb{R},~~|{\rm Re}(\lambda_{i})|=|{\rm Im}(\lambda_{i})|\\
&\epsilon< \min_{\lambda_{i}\neq 0}\frac{{\rm Re}(\lambda_{i})}{{\rm Re}^{2}(\lambda_{i})-{\rm Im}^{2}(\lambda_{i})},~~|{\rm Re}(\lambda_{i})|>|{\rm Im}(\lambda_{i})|\\
&\epsilon> \max_{\lambda_{i}\neq 0}-\frac{{\rm Re}(\lambda_{i})}{{\rm Im}^{2}(\lambda_{i})-{\rm Re}^{2}(\lambda_{i})},~~|{\rm Re}(\lambda_{i})|<|{\rm Im}(\lambda_{i})|
\end{aligned}\right.
\end{equation}
where $ \lambda_{i}$ is the $i$th eigenvalue of matrix $\mathcal {L}$. Moreover, if $\epsilon$ is further required to be positive, then we have
\begin{equation*}
\begin{aligned}
\epsilon\in\bigg(0, ~~\min_{\lambda_{i}\neq 0,~|{\rm Re}(\lambda_{i})|>|{\rm Im}(\lambda_{i})|}\frac{{\rm Re}(\lambda_{i})}{{\rm Re}^{2}(\lambda_{i})-{\rm Im}^{2}(\lambda_{i})}\bigg)
\end{aligned}
\end{equation*}
Moreover, the final aggregated value is restricted in a convex hull spanned by the initial opinions of the roots.
\end{cor}

\begin{IEEEproof}
Let us first study an auxiliary system of $(\ref{20191eq6})$ by
\begin{equation}\label{20191eq7}
\begin{aligned}
\dot{\xi}(t)=&~-W\xi(t),~t\in \mathbb{R}_{+}
\end{aligned}
\end{equation}
where $W=\mathcal {L}-\epsilon\mathcal {L}^{2}$. It is known that system $(\ref{20191eq7})$ guarantees the consensus if and only if matrix $W$ contains a simple zero eigenvalue and the remaining eigenvalues share positive real parts. When the induced graph by Laplacian matrix $\mathcal {L}$ has a directed spanning tree, one has that the zero eigenvalue of matrix $W$ is simple. We continuous to show that the nonzero eigenvalues of $W$ have positive real parts. One can easily find that the nonzero eigenvalues of $W$ are of the form
\begin{equation}\label{20191eq8}
\begin{aligned}
\lambda^{\star}_{i}=& ~{\rm Re}(\lambda_{i})-\epsilon ({\rm Re}^{2}(\lambda_{i})-{\rm Im}^{2}(\lambda_{i}))\\
&+\mathbbm{i}({\rm Im}(\lambda_{i})-2\epsilon{\rm Re}(\lambda_{i}){\rm Im}(\lambda_{i}))
\end{aligned}
\end{equation}
By $(\ref{20191eq8})$, the nonzero eigenvalues in $W$ share positive real parts if and only if $(\ref{ch7_20191eq10_1})$ is desirable.

It is noticeable that the relationship between the eigenvalues of system matrix in $(\ref{20191eq7})$ and those in $(\ref{20191eq6})$ can be formulated by
\begin{equation*}\label{20191eq9}
\begin{aligned}
\lambda^{\ast}_{i}=&~ 1-\varrho\lambda^{\star}_{i},~i\in \mathbb{I}_{N}
\end{aligned}
\end{equation*}
where $\lambda^{\ast}_{i}$ represents the $i$th eigenvalue of $I-\varrho\mathcal {L}+\epsilon\varrho\mathcal {L}^{2}$. As a result, one can check that $I-\varrho\mathcal {L}+\epsilon\varrho\mathcal {L}^{2}$ has a simple $1$ eigenvalue and the remaining eigenvalues are restricted in the unit disk if and only if $(\ref{20191eq10})$ is desirable, by which $| \lambda^{\ast}_{i} |<1$ is always guaranteed as long as $\lambda^{\star}_{i}\neq0$.

The second part is trivial, and hence is omitted.
\end{IEEEproof}

An extension of $\epsilon=\varrho$ would be interesting since the developed method in Corollary $\ref{20191cor1}$ does not work in such a case. Actually, $\varrho$ is involved in both $(\ref{20191eq6})$ and $(\ref{20191eq7})$, leading to the failure of computing the allowable range for $\varrho$. Apart from these concerns, the argument induced within a general interacting network is far from obvious in contrast to the case of bidirectional interacting network, as suggested by Corollary $\ref{20191cor1}$.

\begin{thm}\label{20191th6}
Suppose that the graph induced by $\mathcal {L}$ has a directed spanning tree. Then system $(\ref{20191eq6})$ achieves the \emph{consensus in opinions} if and only if the step size $\varrho$ is bounded with the following constraints,
\begin{itemize}
\item[({\rm \lowercase \expandafter {\romannumeral 1}})] If ${\rm Im}(\lambda_{i})=0$ for $\lambda_{i}\neq0$,
\begin{equation*}\label{20191eq95}
\begin{aligned}
\varrho\in \min_{\lambda_{i}}\bigg(0,\frac{1}{\lambda_{i}}\bigg)
\end{aligned}
\end{equation*}
 where $\lambda_{i}>0$ is the eigenvalue of $\mathcal {L}$.
\item[({\rm \lowercase \expandafter {\romannumeral 2}})] If ${\rm Im}(\lambda_{i})\neq0$ with $\lambda_{i}\neq0$,
\begin{equation*}\label{20191eq66}
\left\{\begin{aligned}
&\min_{\varrho>0,\lambda_{i}\neq0} f_{i}(\varrho,\lambda_{i},\arg(\lambda_{i}))>0\\
&\varrho\not\in\bigg\{\varrho_{i,1},\varrho_{i,2}\bigg\}\bigcup \bigg\{\varrho_{i,3},\varrho_{i,4}\bigg\}
\end{aligned}\right.
\end{equation*}
where
\begin{equation*}\label{20191eq67}
\begin{aligned}
&f_{i}(\varrho,\lambda_{i},\arg(\lambda_{i}))\\
=&~-\varrho^{3} |\lambda_{i}|^{3}+\varrho^{2} |\lambda_{i}|^{2} \cos^{2}(\arg(\lambda_{i}))\\
&-2\varrho |\lambda_{i}|\sin(2\arg(\lambda_{i})) -\varrho |\lambda_{i}|+2\cos(\arg(\lambda_{i}))
\end{aligned}
\end{equation*}
and $\varrho_{i,1}$, $\varrho_{i,2}$, $\varrho_{i,3}$, $\varrho_{i,4}$ are depicted in $(\ref{20191eq107})$
\begin{figure*}
\begin{multicols}{2}
\end{multicols}
\leftsep
\begin{equation}\label{20191eq107}
\left\{\begin{aligned}
\varrho_{i,1}=&~\frac{\cos(\arg(\lambda_{i}))+\sqrt{\cos^{2}(\arg(\lambda_{i}))(8\cos(\theta)-7)+4(1-\cos(\theta))}}{2|\lambda_{i}| \cos(2\arg(\lambda_{i}))    }\\
\varrho_{i,2}=&~\frac{\cos(\arg(\lambda_{i}))-\sqrt{\cos^{2}(\arg(\lambda_{i}))(8\cos(\theta)-7)+4(1-\cos(\theta))}}{2|\lambda_{i}| \cos(2\arg(\lambda_{i}))    }\\
\varrho_{i,3}=&~\frac{\sin(\arg(\lambda_{i}))+\sqrt{\cos^{2}(\arg(\lambda_{i}))(8\cos(\theta)-7)+4(1-\cos(\theta))}}{2|\lambda_{i}| \sin(2\arg(\lambda_{i}))    }\\
\varrho_{i,4}=&~\frac{\sin(\arg(\lambda_{i}))-\sqrt{\cos^{2}(\arg(\lambda_{i}))(8\cos(\theta)-7)+4(1-\cos(\theta))}}{2|\lambda_{i}| \sin(2\arg(\lambda_{i}))    }
\end{aligned}\right.
\end{equation}
where $\theta\in[0, 2\pi)$.
\rightsep
\begin{multicols}{2}
\end{multicols}
\end{figure*}
\end{itemize}
\end{thm}

\begin{IEEEproof}
The proof of Theorem $\ref{20191th6}$ is self-contained, and is reported in
{\scshape Appendix} for the sake of concinnity.
\end{IEEEproof}

From Theorem $\ref{20191th6}$, the requirement on $\varrho$ coincides with the statement in \cite{olfati2004consensus} if the underlying network is bidirectional. For the general interpersonal network, the condition on $\varrho$ is far from trivial as the previous case since it tightly links to the amplitude and the argument principal value of nonzero eigenvalues corresponding to $\mathcal {L}$. In fact, there is an alternative to guarantee the consensus in opinion for cooperative antagonistic network, even though it falls short of elegance as Theorem $\ref{20191th6}$.

\begin{cor}\label{20191cor2}
Suppose that the graph induced by $\mathcal {L}$ attains a directed spanning tree. Then system $(\ref{20191eq6})$ achieves the \emph{consensus in opinions} if and only if
\begin{equation}\label{20191eq63}
\begin{aligned}
\varrho\in\bigcap_{\varrho>0,\lambda_{i}\neq0}\bigg\{
a_{i}\varrho^{3}+b_{i}\varrho^{2}+c_{i}\varrho+d_{i}<0
\bigg\}
\end{aligned}
\end{equation}
where
\begin{equation*}\label{20191eq60}
	\begin{aligned}
		a_{i}=&~4{\rm Re}^{2}(\lambda_{i})+({\rm Re}^{2}(\lambda_{i})-{\rm Im}^{2}(\lambda_{i}))^{2}\\
		b_{i}=&~2{\rm Re}(\lambda_{i})({\rm Re}^{2}(\lambda_{i})-{\rm Im}^{2}(\lambda_{i}))-4{\rm Re}(\lambda_{i}){\rm Im}(\lambda_{i})\\
		c_{i}=&~3{\rm Re}^{2}(\lambda_{i})-{\rm Im}^{2}(\lambda_{i})\\
		d_{i}=&~-2{\rm Re}(\lambda_{i})
\end{aligned}
\end{equation*}
In addition, consensus in system $(\ref{20191eq6})$ is achieved for the undirected graph if and only if
\begin{equation*}\label{20191eq60_1}
\begin{aligned}
\varrho\in \min_{\lambda_{i}}\bigg(0,\frac{1}{\lambda_{i}}\bigg)
\end{aligned}
\end{equation*}
 where $\lambda_{i}>0$ is the eigenvalue of $\mathcal {L}$.
\end{cor}

\begin{IEEEproof}
System $(\ref{20191eq6})$ can be rewritten as
\begin{equation}\label{20191eq64}
\begin{aligned}
\xi(k+1)=&~(I-\varrho\mathcal {L}+\varrho^{2}\mathcal {L}^{2})\xi(k)
\end{aligned}
\end{equation}
The eigenvalue of matrix $I-\varrho\mathcal {L}+\varrho^{2}\mathcal {L}^{2}$ is of the form
\begin{equation*}\label{20191eq61}
\begin{aligned}
\lambda^{\ast}_{i}=&~ 1-\varrho\lambda_{i}+\varrho^{2}\lambda^{2}_{i}\\
=&~ 1-{\rm Re}(\lambda_{i})\varrho+ \bigg({\rm Re}^{2}(\lambda_{i})-{\rm Im}^{2}(\lambda_{i})\bigg)\varrho^{2}\\
&+\bigg(2{\rm Re}(\lambda_{i}){\rm Im}(\lambda_{i})\varrho^{2}     -{\rm Im}(\lambda_{i})\varrho\bigg)\mathbbm{i}
\end{aligned}
\end{equation*}
Therefore, $|\lambda^{\ast}_{i}|<1$ with ${\rm Re}(\lambda_{i})>0$ is equal to
\begin{equation*}
\begin{aligned}
1\geq&~\bigg(1-{\rm Re}(\lambda_{i})\varrho+ \bigg({\rm Re}^{2}(\lambda_{i})-{\rm Im}^{2}(\lambda_{i})\bigg)\varrho^{2}\bigg)^{2}\\
&+\bigg(2{\rm Re}(\lambda_{i}){\rm Im}(\lambda_{i})\varrho^{2}     -{\rm Im}(\lambda_{i})\varrho\bigg)^{2}
\end{aligned}
\end{equation*}
By tedious computation, it yields the requirement in $(\ref{20191eq63})$. And the second statement is consistent with that in Theorem $\ref{20191th6}$, and is hence omitted.
\end{IEEEproof}


\begin{figure}
\centering
\includegraphics[width=3.45in,height=1.1in]{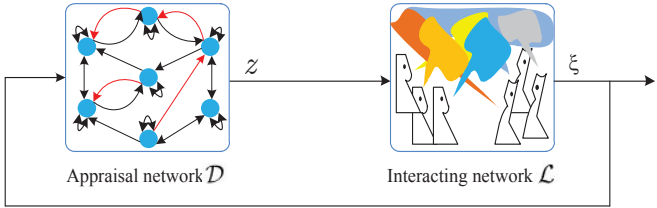}
\caption{A holistic paradigm of the opinion evolution for system $(\ref{20191eq2})$ where the appraisal network (characterized by $\mathcal {D}$) is antagonistic, that is, the red arcs mean $\delta_{ij}<0$ while the black ones indicate that $\delta_{ij}>0$.}
\label{20191_fig1_2}
\end{figure}

\subsection{Antagonistic Appraisal Network}\label{20191 sec42}
A remarkable feature of the social networks is that social actors rarely share unanimous opinions. It has been recognized that biased assimilation principle or                                                                                                                                                                                                                                                                                                                                                                                                                                                                                                                                                                                                                                                                                                                                                                                                                                                                                                                                                                                                                                                                                                                                                                                                                                                                                                                                                                                                                                                                                                                                                                                                                                                                                                                                                                                                                                                                                                                                                                                                                                                                                                                                                                                                                                                                                                                                                                                                                                                                                                                                                                                                                                                                                                                                                                                                                                                                                                                                                                                                                                                                                                                                                                                                                                                                                                                                                                                                                                                                                                                                                                                                                                                                                                                                                                                                                                                                                                                                                                                                                                                                                                                                                                                                                                                                                                                                                                                                                                                                                                                                                                                                                                                                                                                                                                                                                                                                                                                                                                                                                                                                                                                                                                                                                                                                                                                                                                                                                                                                                                                                                                                                                                                                                                                                                                                                                                                                                                                                                                                                                                                                                                                                                                                                                                                                                                                                                                                                                                                                                                                                                                                                                                                                                                                                                                                                                                                                                                                                                                                                                                                                                                                                                                                                                                                                                                                                                                                                                                                                                                                                                                                                                                                                                                                                                                                                                                                                                                                                                                                                                                                                                                                                                                                                                                                                                                                                                                                                                                                                                                                                                                                                                                                                                                                                                                                                                                                                                                                                                                                                                                                                                                                                                                                                                                                                                                                                                                                                                                                                                                                                                                                                                                                                                                                                                                                                                                                                                                                                                                                                                                                                                                                                                                                                                                                                                                                                                                                                                                                                                                                                                                                                                                                                                                                                                                                                                                                                                                                                                                                                                                                                                                                                                                                                                                                                                                                                                                                                                                                                                                                                                                                                                                                                                                                                                                                                                                                                                                                                                                                                                                                                                                                                                                                                                                                                                                                                                                                                                                                                                                                                                                                                                                                                                                                                                                                                                                                                                                                                                                                                                                                                                                                                                                                                                                                                                                                                                                                                                                                                                                                                                                                                                                                                                                                                                                                                                                                                                                                                                                                                                                                                                                                                                                                                                                                                                                                                                                                                                                                                                                                                                                                                                                                                                                                                                                                                                                                                                                                                                                                                                                                                                                                                                                                                                                                                                                                                                                                                                                                                                                                                                                                                                                                                                                                                                                                                                                                                                                                                                                                                                                                                                                                                                                                                                                                                                                                                                                                                                                                                                                                                                                                                                                                                                                                                                                                                                                                                                                                                                                                                                                                                                                                                                                                                                                                                                                                                                                                                                                                                                                                                                                                                                                                                                                                                                                                                                                                                                                                                                                                                                                                                                                                                                                                                                                                                                                                                                                                                                                                                                                                                                                                                                                                                                                                                                                                                                                                                                                                                                                                                                                                                                                                                                                                                                                                                                                                                                                                                                                                                                                                                                                                                                                                                                                                                                                                                                                                                                                                                                                                                                                                                                                                                                                                                                                                                                                                                                                                                                                                                                                                                                                                                                                                                                                                                                                                                                                                                                                                                                                                                                                                                                                                                                                                                                                                                                                                                                                                                                                                                                                                                                                                                                                                                                                                                                                                                                                                                                                                                                                                                                                                                                                                                                                                                                                                                                                                                                                                                                                                                                                                                                                                                                                                                                                                                                                                                                                                                                                                                                                                                                                                                                                                                                                                                                                                                                                                                                                                                                                                                                                                                                                                                                                                                                                                                                                                                                                                                                                                                                                                                                                                                                                                                                                                                                                                                                                                                                                                                                                                                                                                                                                                                                                                                                                                                                                                                                                                                                                                                                                                                                                                                                                                                                                                                                                                                                                                                                                                                                                                                                                                                                                                                                                                                                                                                                                                                                                                                                                                                                                                                                                                                                                                                                                                                                                                                                                                                                                                                                                                                                                                                                                                                                                                                                                                                                                                                                                                                                                                                                                                                                                                                                                                                                                                                                                                                                                                                                                                                                                                                                                                                                                                                                                                                                                                                                                                                                                                                                                                                                                                                                                                                                                                                                                                                                                                                                                                                                                                                                                                                                                                                                                                                                                                                                                                                                                                                                                                                                                                                                                                                                                                                                                                                                                                                                                                                                                                                                                                                                                                                                                                                                                                                                                                                                                                                                                                                                                                                                                                                                                                                                                                                                                                                                                                                                                                                                                                                                                                                                                                                                                                                                                                                                                                                                                                                                                                                                                                                                                                                                                                                                                                                                                                                                                                                                                                                                                                                                                                                                                                                                                                                                                                                                                                                                                                                                                                                                                                                 homophily principle (cf. \cite{dandekar2013biased}) is vital to explain the opinion clusters. Another reason for the opinion dynamics giving rise to clusters is antagonism. The main goal of this subsection is to investigate the case where there exists antagonistic interactions among social actors.

By following the basic route as that in subsection III-A, the public network characterized by $\mathcal{L}$ is the same as before. Meanwhile, the appraisal network characterized by $\mathcal {D}$ involves hostile interactions. A simple illustration for system $(\ref{20191eq2})$ is depicted by Fig. $\ref{20191_fig1_2}$.

Suppose that agent $i$ possesses an opposite attitude to the opinion of agent $j$, then ${\rm sgn}(\delta_{ij})=-1$. In addition, since $|\delta_{ij}|$ quantifies the degree of the influence among the total social influence that the $i$th social actor has, we require $\sum^{N}_{j=1}|\delta_{ij}|=1$ for all $i$. Before moving on, a condiment is needed.

\begin{de}\label{20191de2}
Two graphs $\mathcal{G}_{1}$ and $\mathcal{G}_{2}$ share the same topology if there exists an edge from the $i$th node to the $j$th node in $\mathcal{G}_{1}$, then the edge $(i,j)$ also belongs to  $\mathcal{G}_{2}$, and vice versa.
\end{de}

Actually, two graphs satisfying Definition $\ref{20191de2}$ differ only in the weighted values of edges. Therefore, we can define a new stochastic matrix $\mathcal {D}^{\star}\triangleq (|\delta_{ij}|)_{N\times N}$. With the help of Definition $\ref{20191de2}$, the graphs induced by $\mathcal {D}$ and $\mathcal {D}^{\star}$ have a topology structure in common.
\begin{thm}\label{20191th2}
Suppose that the appraisal network is antagonistic. Then system $(\ref{20191eq2})$ \emph{converges in opinions} if and only if matrix $I-\Lambda\mathcal {L}\mathcal {D}$ has a simple $1$ eigenvalue and the remaining eigenvalues are preserved in the unit disk.
\end{thm}


\begin{IEEEproof}
The proof of Theorem $\ref{20191th2}$ is similar to that in Theorem $\ref{20191th1}$. Generally speaking, the right eigenvector $\iota$ associated with the zero eigenvalue of $\Lambda\mathcal {L}\mathcal {D}$ may not be equal to vector one when the appraisal network contains hostile interactions. Therefore, it follows that $\varphi=\iota\varsigma^{\prime}\xi(0)$. This implies the convergence in opinions of the social actors.
\end{IEEEproof}

According to Theorem $\ref{20191th2}$, the clusters in opinions occur due to the presence of antagonistic interactions in the appraisal network. In view of Theorems $\ref{20191th1}$ and $\ref{20191th2}$, it can be concluded that cooperative appraisal networks lead to the consensus in opinions, while hostile appraisal networks attain the clusters in opinions. Here we emphasize that it is rather arduous to give some specific indices on the weighted influence matrix $\Lambda$ due to the potential complexity of the appraisal network. Another interesting welfare behind Theorems $\ref{20191th1}$ and $\ref{20191th2}$ is that the agents in system $(\ref{20191eq2})$ will not converge to zero for almost all initial values. This is because that $I-\Lambda\mathcal {L}\mathcal {D}$ contains at least an eigenvalue one. Fortunately, this intriguing feature could be addressed by virtue of the issue dependence matrix that is about to be elucidated later.


\begin{figure}
\centering
\includegraphics[width=3.75in,height=2.5in]{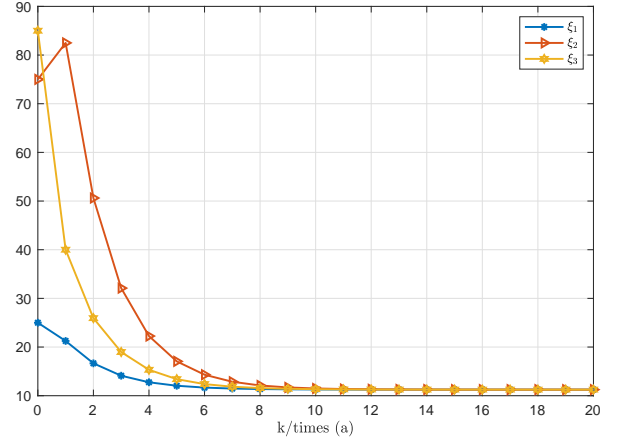}\\
\includegraphics[width=3.75in,height=2.5in]{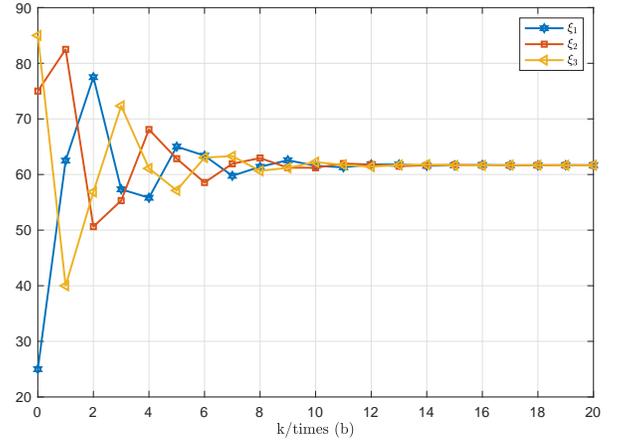}
\caption{(a) Opinion evolution for the antagonistic appraisal network where the initial opinions $\xi(0)=(25,75,85)$ are partly borrowed from \cite[Equation (15)]{parsegov2017novel}; (b) Opinion evolution for the antagonistic appraisal network with the same parameters as (a) except for replacing $\Lambda$ by $0.5I$.}
\label{20191_fig2}
\end{figure}
\subsection{Connection between Cooperative and Antagonistic Appraisal Networks}\label{20191 sec44}
Subsections ${\rm \ref{20191 sec41}}$ and ${\rm \ref{20191 sec42}}$ discuss the opinion evolution on cooperative and antagonistic appraisal networks, respectively. Here we are interested in a question: what is the bridge between the cooperative and the antagonistic appraisal networks? To answer this question, we first look at a simple example.

\begin{exa}\label{20191exa1}
Consider a network with three social actors whose interacting and appraisal networks are of the forms
\begin{equation*}\label{20191eq45}
\begin{aligned}
\mathcal{L}=\begin{bmatrix}
2& -1 &-1\\
-1& 2 &-1\\
-1& -1 &2
\end{bmatrix},~\mathcal {D}=\begin{bmatrix}
0.5& -0.5& 0\\
0& 0.5& -0.5\\
-0.5& 0& 0.5
\end{bmatrix}
\end{aligned}
\end{equation*}
We choose $\Lambda={\rm diag}(-0.05, 0.5,0.5)$, and then compute the eigenvalues of $I-\Lambda\mathcal {L}\mathcal {D}$
\begin{equation*}\label{20191eq46}
\begin{aligned}
\lambda(I-\Lambda\mathcal {L}\mathcal {D})\in \bigg\{0,0.9448,1.9052\bigg\}
\end{aligned}
\end{equation*}
Opinion evolutions of the social actors are plotted in Fig. $\ref{20191_fig2}$(a). It reveals an interesting phenomenon that antagonistic appraisal networks could achieve consensus in opinions. More importantly, we can see that the final aggregated opinion is not restricted in the convex hull spanned by the initial opinions. Next, we select another group of susceptibility factors by $0.5 I$. In this case, the consensus in opinions could still be preserved while it is contained in the convex hull spanned by the initial opinions (see Fig. $\ref{20191_fig2}$(b) for more details). Therefore, we could treat the susceptibility factor $\varrho_{i}$ as a design parameter quantifying how does the opinion formation process work. \QEDA
\end{exa}

Note that consensus in opinions could be preserved for cooperative appraisal network. It implicitly manifests that the opinions either achieve consensus or diverge whenever the appraisal network is cooperative. However, opinions either achieve consensus or clusters if the underlying appraisal network is antagonistic, except for the divergence in opinions. We summarize them as below.

\begin{pro}\label{20191pro2}
Suppose that the appraisal network is antagonistic. Then system $(\ref{20191eq2})$ achieves \emph{consensus in opinions} if and only if the following properties hold:
\begin{itemize}
\item[({\rm \lowercase \expandafter {\romannumeral 1}})] System $(\ref{20191eq2})$ converges;
\item[({\rm \lowercase \expandafter {\romannumeral 2}})] $\mathcal {D}\textbf{1}_{N}=\mathcal {O}_{N\times1}$ or $\mathcal {D}\textbf{1}_{N}=-\textbf{1}_{N}$.
\end{itemize}
\end{pro}

\begin{IEEEproof}
(Necessity) According to Definition $\ref{20191de1}$, the consensus in $(\ref{20191eq2})$ indicates that $\lim_{k\rightarrow\infty}\xi_{i}(k)=\lim_{k\rightarrow\infty}\xi_{j}(k)$ for arbitrary $i,j\in \mathbb{I}_{N}$. Hence, $\textbf{1}_{N}$ is the right eigenvector associated with the eigenvalue one of matrix $I-\Lambda\mathcal {L}\mathcal {D}$. It further arrives at
\begin{equation*}\label{20191eq47}
\begin{aligned}
\mathcal {O}_{N\times1}=&\Lambda\mathcal {L}\mathcal {D}\textbf{1}_{N}\Rightarrow \mathcal {O}_{N\times1}=\mathcal {L}\mathcal {D}\textbf{1}_{N}
\end{aligned}
\end{equation*}
The above formula is true since matrix $\Lambda$ is nonsingular. The directed spanning tree preserved in the interacting network implies that merely one of the following equalities is satisfied,
\begin{subequations}\label{20191eq48}
\begin{empheq}[left=\empheqlbrace]{align}
\mathcal {D}\textbf{1}_{N}= &~\textbf{1}_{N} \label{20191eq48a}\\
\mathcal {D}\textbf{1}_{N}=&~\mathcal {O}_{N\times1}  \label{20191eq48b}\\
\mathcal {D}\textbf{1}_{N}=&~-\textbf{1}_{N}  \label{20191eq48c}
\end{empheq}
\end{subequations}
Next we will show that only $(\ref{20191eq48b})$ or $(\ref{20191eq48c})$ hold for the antagonistic appraisal network. Suppose that $(\ref{20191eq48a})$ holds, i.e.,
\begin{equation}\label{20191eq49}
\begin{aligned}
\sum^{N}_{j=1}\delta_{ij}=1, ~\forall~i\in \mathbb{I}_{N}
\end{aligned}
\end{equation}
Moreover, we also require
\begin{equation}\label{20191eq50}
\begin{aligned}
\sum^{N}_{j=1}|\delta_{ij}|=1, ~\forall~i\in \mathbb{I}_{N}
\end{aligned}
\end{equation}
One can see that $(\ref{20191eq49})$ and $(\ref{20191eq50})$ suggest
\begin{equation*}\label{20191eq51}
\begin{aligned}
|\delta_{ij}|=\delta_{ij}, ~\forall~i,j\in \mathbb{I}_{N}
\end{aligned}
\end{equation*}
which implies that the appraisal network is cooperative. It is a contradiction. Therefore, we always have $\mathcal {D}\textbf{1}_{N}=\mathcal {O}_{N\times1}$ or $\mathcal {D}\textbf{1}_{N}=-\textbf{1}_{N}$. Obviously, $\mathcal {D}\textbf{1}_{N}=-\textbf{1}_{N}$ means that all entries of $\mathcal {D}$ are non-positive, which indicates that the appraisal network is antagonistic.

(Sufficiency) The convergence of system $(\ref{20191eq2})$ means that $I-\Lambda\mathcal {L}\mathcal {D}$ has a simple eigenvalue one (note that system $(\ref{20191eq2})$ cannot guarantee the stability since matrix $\Lambda\mathcal {L}\mathcal {D}$ always has at least a zero eigenvalue). In addition, $\mathcal {D}\textbf{1}_{N}=\mathcal {O}_{N1}$ or $\mathcal {D}\textbf{1}_{N}=-\textbf{1}_{N}$ suggests that the vector $\textbf{1}_{N}$ is an eigenvector associated with the eigenvalue one of matrix $I-\Lambda\mathcal {L}\mathcal {D}$. Therefore, system $(\ref{20191eq2})$ preserves consensus in opinions.
\end{IEEEproof}

With the help of Proposition $\ref{20191pro2}$ and Theorem $\ref{20191th2}$, we could conclude that cooperative appraisal network achieves consensus in opinions, while antagonistic appraisal network shows both consensus and clusters in opinions. However, antagonistic appraisal network that can achieve consensus in opinions has certain special requirements for its structure.

\subsection{Estimation for Appraisal Network}\label{20191 sec43}
Briefly speaking, it is a thorny problem to determine the structure of the appraisal network. Fortunately, a large number of efforts have been poured on identifying the dynamic structure and topologies of the networks with the help of the experiment data. The emerging fields include sociology, signal processing and statistics (see, e.g., \cite{friedkin2011social,snijders2010maximum,wai2016active}). As hinted before, interacting network is always known to all; while appraisal network is generally private, and thus is of great significance to be estimated. The reason lies that it is the first step to understand the mechanism behind the emerging and the evolution of the opinions in social networks.

To cope with the estimation issue related to appraisal network, a technical lemma is needed.
 \begin{lem}(\hspace{-0.001cm}\cite{laub2005matrix})\label{20191le3}
 	For any matrix $Q=(q_{1},...,q_{n})$, its vectorization, denoted by ${\rm vec}(Q)$, is
 	\begin{equation*}
 	\begin{aligned}
 	{\rm vec}(Q)=(q^{\prime}_{1},...,q^{\prime}_{n})^{\prime}
 	\end{aligned}
 	\end{equation*}
 	For matrices $Q$, $W$ and $R$, the vectorization with respect to their product is given by
 	\begin{equation*}\label{20191eq32}
 	\begin{aligned}
 	{\rm vec}(QWR)=& (R^{\prime}\otimes Q){\rm vec}(W)
 	\end{aligned}
 	\end{equation*}
 \end{lem}

By virtue of Lemma $\ref{20191le3}$, two properties can be obtained.
\begin{cor}\label{20191cor4}
For any variable $\xi\in\mathbb{R}^{N}$ and matrix $\Lambda\mathcal {L}\mathcal {D}$, the following facts are true
\begin{equation*}
\left\{\begin{aligned}
&{\rm vec}(\xi)= ~\xi\\
&{\rm vec}(\Lambda\mathcal {L}\mathcal {D}\xi)=~(\xi^{\prime}\otimes \Lambda\mathcal {L}){\rm vec}(\mathcal {D})
\end{aligned}\right.
\end{equation*}
\end{cor}

\begin{IEEEproof}
The proof is trivial by Lemma $\ref{20191le3}$.
\end{IEEEproof}

With the above preparations, we are about to formulate the estimation problem on appraisal network. Apart from $\Lambda$ and $\mathcal {L}$, we postulate that one has access to $m$ observed opinions of length $1$ during opinion formation process under $(\ref{20191eq1})$, i.e., $m$ sequences of opinions $(\xi_{t}(k-1),\xi_{t}(k))$ with $t\in\mathbb{I}_{m}$ and $k\in \mathbb{Z}_{+}$, in a manner of independent and identically distribution (i.i.d.). More specifically, we associate a uniform observation of $m$ by
\begin{equation}\label{20191eqnew1}
\begin{aligned}
\Omega_{m}=~\bigg\{   (\xi_{t}(k-1),\xi_{t}(k)), 1\leq t \leq m, \forall~k \in \mathbb{Z}_{+} \bigg\}
\end{aligned}
\end{equation}

Upon collecting the $m$ (randomly with uniform) opinion sets and using Corollary $\ref{20191cor4}$, we formulate the following optimization problem:
\begin{equation}\label{20191eq31}
\begin{aligned}
&~~~~~~~~~~~~~~~~~\min_{\zeta,~Q}~\gamma \\
&{\rm s.t.}~~~~~~f(\zeta,m)=\frac{1}{m}\sum^{m}_{t=1} g(t)\leq \gamma\\
&g(t)=X^{\prime}_{t}(k)QX_{t}(k)\\
&X_{t}(k)=\xi_{t}(k)-\xi_{t}(k-1)+\bigg(\xi^{\prime}_{t}(k-1)\otimes \Lambda\mathcal {L}\bigg)\zeta\\
&\gamma \geq0, ~Q\succ 0
\end{aligned}
\end{equation}

\begin{rmk}
For optimization problem $(\ref{20191eq31})$, we emphasize: ({\rm \lowercase \expandafter {\romannumeral 1}}) The utilization of Corollary $\ref{20191cor4}$ is helpful since  the coupling between the interacting network and the appraisal network is entangled; ({\rm \lowercase \expandafter {\romannumeral 2}}) In general, it is unrealistic to require parameter $m$ to be infinite because system matrix in $(\ref{20191eq2})$ is not stable.
\end{rmk}

Algorithm $\ref{20191alg1}$ depicted below gives a lower bound on the ``observations" needed for a desirable performance.

\begin{algorithm}[!h]
	\caption{Lower Bound for the Observations in $\Omega_{m}$}\label{20191alg1}
	\begin{algorithmic}[1]
		\STATE Given a sequence of observations  of length $m$, and a prior level of performance index $0<\gamma_{0}<\infty$;
		\STATE Set  $m=m_{0}$ with $m_{0}\geq 1$;
		\WHILE{$\gamma>\gamma_{0}$;}
		\STATE Print: Gauge of appraisal network fails;
		\STATE Set $m\leftarrow1+m$;
		\STATE Optimize $(\ref{20191eq31})$;
		\ENDWHILE
		\RETURN $m$;
	\end{algorithmic}
\end{algorithm}

Although Algorithm 1 may provide an estimation on appraisal network, we still cannot give specific value of sample length $m$ that is essential for solving $(\ref{20191eq31})$. Obviously, too much observations bring computation complexity, while less observation may yield inaccuracy estimation. Furthermore, even Algorithm $\ref{20191alg1}$ may mitigate the downside on $(\ref{20191eq31})$ to a certain degree, a lot of computing resources are needed apart from the time consumption since Algorithm $\ref{20191alg1}$ executes in a way of try-once-discard. All the discussions above naturally raise an interesting question: What is the confidence can we possess with the aid of a finite sets on opinions of empirical observations, on which the appraisal network can work for the whole opinion space? To this end, we resort to the tool from random convex optimization (also called by chance-constrained optimization); see Ref. \cite{calafiore2006scenario,calafiore2010random} for more details.

Now, we are going to formulate the considered problem. For fixed probability space $(\Omega, \mathcal{F}, \mathcal{P})$, compact set $\mathbb{X}\subseteq \mathbb{R}^{N\times N}$ is convex and contains the origin as its interior. A measurable function $f: \mathbb{X}\times \Omega\mapsto\mathbb{R}$ is convex on its first argument for any fixed second argument, and bounded for the second argument whenever the first is fixed. Thereby, for random finite opinion set $\Omega_{m}$ in $(\ref{20191eqnew1})$ and any level parameter $\varepsilon\in (0,1)$, given a confidence level $\beta\in (0,1)$ and some symmetric positive definite matrix $Q$, the probability of violation (cf. \cite[Definition 1]{calafiore2006scenario}) is
\begin{equation*}
\begin{aligned}
V(\zeta,m)=\mathcal{P}\{m\in \mathbb{Z}_{+}: f(\zeta,m)>\gamma^{\star}\}
\end{aligned}
\end{equation*}
where $\gamma^{\star}$ is the optimal solution of the optimization problem $(\ref{20191eq31})$.

We then focus on the following optimization problem:
\begin{equation}\label{20191eqnew2}
\begin{aligned}
\min_{\zeta}~~~~~~&c^{\prime} \zeta\\
{\rm s.t.}~~~~~~&\mathcal{P}\{(\zeta,m): V(\zeta,m)> \varepsilon\}\leq \beta\\
&\zeta \in \mathbb{X}
\end{aligned}
\end{equation}
 where $c$ is a certain ``cost" vector (also known as the objective direction).

For optimization problem $(\ref{20191eqnew2})$, we get the probability with respect to $f(\zeta,m)\leq\gamma^{\star}$ at least $1-\beta$ using the samples with length $m$. A smaller violation level $\varepsilon$ is a more desirable estimation on the appraisal network. As a result, the number of samples increases. The next theorem gives a confirmative answer on how many opinion observations are needed to obtain a satisfactory performance.

 \begin{thm}\label{20191th5}
Consider optimization problem $(\ref{20191eqnew2})$ with $m\geq N^{2}$ where $\Omega_{m}\subseteq\mathbb{X}$, in the sense of i.i.d. Then, for $\forall\varepsilon \in[0,1]$, it follows that
\begin{equation*}
\begin{aligned}
\mathcal{P}\{(\zeta,m): V(\zeta,m)> \varepsilon\}\leq \beta(\varepsilon,m)
\end{aligned}
\end{equation*}
where
\begin{equation*}
\begin{aligned}
\beta(\varepsilon,m)=\sum^{m}_{\ell=0}
\begin{pmatrix}
N^{2}\\
\ell
\end{pmatrix}\varepsilon^{\ell}(1-\varepsilon)^{m-\ell}
\end{aligned}
\end{equation*}
Moreover, the low bound on $m$ is
\begin{equation*}
\begin{aligned}
m(\varepsilon,\beta)=\min\bigg\{  m\in \mathbb{Z}_{+}\bigg|  \sum^{m}_{\ell=0}
\begin{pmatrix}
N^{2}\\
\ell
\end{pmatrix}\varepsilon^{\ell}(1-\varepsilon)^{m-\ell}\leq \beta  \bigg\}
\end{aligned}
\end{equation*}
 \end{thm}

\begin{IEEEproof}
By \cite[Theorem 3.3]{calafiore2010random} and \cite[Theorem 1]{campi2008exact}, the proof follows directly.
\end{IEEEproof}

 \begin{rmk}
Theorem  $\ref{20191th5}$ explicitly provides the number of required samples to obtain a ``good" estimation for the true appraisal network. In \cite{parsegov2017novel}, the authors concentrated on the estimation of multi-issue dependence matrix. We do some further work in this paper. That is, on the one hand, we can give a desirable estimation on the appraisal network. On the other hand, a bound on the samples is provided which has no been addressed in the previous work.
 \end{rmk}

\section{Topic Specific Opinion}\label{20191 sec4}
Unlike man-made systems, the actors in social networks are generally affected by a couple of topics that are interdependent. For example, the leader of the corporate should account for many factors if he/she intends to operate a policy, such as the cost, the external and the internal environment, the potential market as well as the potential customers, etc. As a matter of fact, the issue dependence problem has been well addressed for a long time, in particular the disciplines such as social anthropology, sociology and political science and psychology where they share a common ground that certain objects are coupled by interdependent cognitive orientations.

The first step on how agents' interpersonal influences form a belief system was proposed by the FJ model (cf.  \cite{friedkin1999social})
\begin{equation}\label{20191eq52}
\begin{aligned}
\xi_{i}(k+1)=&\lambda_{ii}C\sum^{N}_{j=1}p_{ij}\xi_{j}(k)+(1-\lambda_{ii})\mu_{i}
\end{aligned}
\end{equation}
where $\xi_{i}(k)\in\mathbb{R}^{n}$, $\lambda_{ii}$ and $\mu_{i}$ are, respectively, the susceptibility and the initial opinion of the $i$th agent. $P=(p_{ij})_{N\times N}$ with $p_{ij}\geq0$ is a stochastic matrix. Matrix $C=(c_{ij})_{n\times n}$ stands for the introspective transformation called the multi-issues dependence structure (MiDS) (cf. \cite{parsegov2017novel}), and satisfies $\sum^{n}_{j=1}|c_{ij}|=1$.

For topic specific issues, the opinion evolving equation $(\ref{20191eq1a})$ becomes
\begin{equation}\label{20191eq53}
\begin{aligned}
\xi_{i}(k+1)=&~C\xi_{i}(k)+\varrho_{i}C\sum_{j\in\mathcal{N}_{i}}  a_{ij}(z_{j}(k)-z_{i}(k))
\end{aligned}
\end{equation}
where the constant matrix $C\in\mathbb{R}^{n\times n}$ describes the MiDS, while the remaining variables are the same as those in $(\ref{20191eq1a})$. Combining $(\ref{20191eq1b})$ and $(\ref{20191eq53})$ gives
\begin{equation}\label{20191eq54}
\begin{aligned}
\xi(k+1)=&~(I-\Lambda\mathcal {L}\mathcal {D})\otimes C\xi(k)
\end{aligned}
\end{equation}
where $\otimes$ denotes the Kronecker product.

Here we discuss the similarities and the differences between $(\ref{20191eq52})$ and $(\ref{20191eq54})$. \\
\textbf{Similarity}:\\
 They both model how the interdependent issues affect the opinion's evolution. \\
 \textbf{Differences:}\\
$(\ref{20191eq52})$ is concerned with the opinion evolution in the context of the cooperative interacting networks while $(\ref{20191eq54})$ studies the opinion evolution with antagonistic interactions characterized by an appraisal network. We emphasize that the interacting network in $(\ref{20191eq54})$ merely quantifies whether or not there exists an information flow between a pair of agents. Moreover, we can further extend $(\ref{20191eq54})$ to the case where some agents may never forget their initial opinions. However, this is beyond the scope of this paper, and is thus omitted.

Note that $(\ref{20191eq2})$ usually ensures the convergence in opinions. However, agents in $(\ref{20191eq54})$ may converge to zero due to the appearance of matrix $C$.
\begin{thm}\label{20191th3}
Suppose that matrix $I-\Lambda\mathcal {L}\mathcal {D}$ has a simple eigenvalue one. The agents in $(\ref{20191eq54})$ achieve the \emph{stability} if and only if $|\lambda_{\max}(C)|<1$.
\end{thm}


\begin{IEEEproof}
It is more vulnerable to check that the system is stable if and only if the eigenvalues in matrix $(I-\Lambda\mathcal {L}\mathcal {D})\otimes C$ are constrained in the unit disk. Note that the eigenvalues of matrix $(I-\Lambda\mathcal {L}\mathcal {D})\otimes C$ are $\lambda(I-\Lambda\mathcal {L}\mathcal {D})\lambda(C)$ according to the matrix theory. Hence, $(\ref{20191eq54})$ is stable if and only if $|\lambda_{\max}(I-\Lambda\mathcal {L}\mathcal {D})\lambda_{\max}(C)|<1$. This is equivalent to $|\lambda_{\max}(C)|<1$. The proof hence follows.
\end{IEEEproof}

Theorem $\ref{20191th3}$ suggests that we can achieve the stability of the agents by just imposing the restriction on matrix $C$ provided that the issue-free cases are convergent. Note that the number of issues are drastically less than that of the participating individuals in general. An interesting welfare from Theorem $\ref{20191th3}$ in contrast to Theorems $\ref{20191th1}$ and $\ref{20191th2}$ is the stability of the interacting agents, which is another motivation to introduce the issue-interdependence matrix $C$ for setup $(\ref{20191eq1})$. Indeed, only the convergence of the opinions is generally assured for the issue-free scenario (see Theorems $\ref{20191th1}$ and $\ref{20191th2}$ for more information).

As discussed before, the social networks rarely achieve unanimous behavior (we treat the stability of the agents by a special case of the consensus). Therefore, it is of great importance to further study the convergence condition on $(\ref{20191eq54})$ in the presence of matrix $C$.
\begin{thm}\label{20191th4}
Suppose that matrix $I-\Lambda\mathcal {L}\mathcal {D}$ has a simple eigenvalue one. The agents in $(\ref{20191eq54})$ are \emph{convergent} if and only if $|\lambda^{\star}_{\max}(I-\Lambda\mathcal {L}\mathcal {D})\lambda_{\max}(C)|<1$ where $\lambda^{\star}_{\max}(I-\Lambda\mathcal {L}\mathcal {D})$ stands for the eigenvalue of $I-\Lambda\mathcal {L}\mathcal {D}$ with the second largest magnitude compared with eigenvalue $1$.
\end{thm}

\begin{IEEEproof}
Here we prove the theorem by following the idea from $(\ref{20191eq11})$ and $(\ref{20191eq54})$. The error system is revised as
\begin{equation}\label{20191eq55}
\begin{aligned}
\theta(k+1)=(I-\Lambda\mathcal {L}\mathcal {D})\otimes C\theta(k)
\end{aligned}
\end{equation}
where $\theta(k) \in \bigg(\mathbb{R}^{N}\backslash \{ \varphi\}\bigg)\otimes \mathbb{R}^{n}$. As $I-\Lambda\mathcal {L}\mathcal {D}$ has a simple eigenvalue one over the space $\theta(k) \in \bigg(\mathbb{R}^{N}\backslash \{ \varphi\}\bigg)\otimes \mathbb{R}^{n}$, the eigenvalues in matrix $I-\Lambda\mathcal {L}\mathcal {D}$ are completely contained in the unit disk. Therefore, we can see that the error system $(\ref{20191eq55})$ is stable if and only if $|\lambda^{\star}_{\max}(I-\Lambda\mathcal {L}\mathcal {D})\lambda_{\max}(C)|<1$. The proof hence follows.
\end{IEEEproof}

Based on the conclusion in Theorem $\ref{20191th4}$, it does not require $|\lambda_{\max}(C)|<1$ as Theorem $\ref{20191th3}$. Hence, there is a matrix $C$ with $|\lambda_{\max}(C)|>1$ that we could still guarantee the convergence of the participating agents as long as $|\lambda^{\star}_{\max}(I-\Lambda\mathcal {L}\mathcal {D})\lambda_{\max}(C)|<1$ is desirable. The following corollary gives some specific requirements on the matrix for guaranteeing the convergence of the agents.

\begin{cor}\label{20191cor3}
Suppose that matrix $I-\Lambda\mathcal {L}\mathcal {D}$ has a simple eigenvalue one. The agents in $(\ref{20191eq54})$ are \emph{convergent} only if $\lim_{k\rightarrow\infty}C^{k}$ exists.
\end{cor}

\begin{IEEEproof}
One can see that
\begin{equation}\label{20191eq56}
\begin{aligned}
\xi(k)=&~(I-\Lambda\mathcal {L}\mathcal {D})^{k}\otimes C^{k}\xi(0)
\end{aligned}
\end{equation}
Therefore, $(\ref{20191eq56})$ converges only if $\lim_{k\rightarrow\infty}C^{k}$ exists. This ends the proof.
\end{IEEEproof}

Due to the antagonistic information, matrix $I-\Lambda\mathcal {L}\mathcal {D}$ may not be a nonnegative stochastic matrix in general. Consequently, the developed methods in the framework of the multi-agent systems are no longer applicable. By virtue of the approach developed in Subsection ${\rm \ref{20191 sec43}}$, it is possible to estimate matrix $C$ if $\Lambda$ and $\mathcal {D}$ are available.

\begin{figure}
	\centering
	\includegraphics[width=1.45in,height=1.5in]{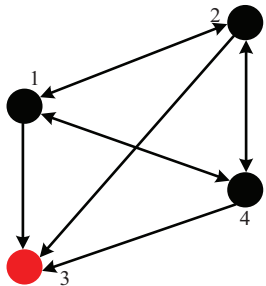}\\
	\caption{Interacting graph of four social actors associated with $(\ref{20191eq44})$.}
	\label{ch7_interacting_graph}
\end{figure}

\begin{figure}
\centering
\includegraphics[width=3.75in,height=2.5in]{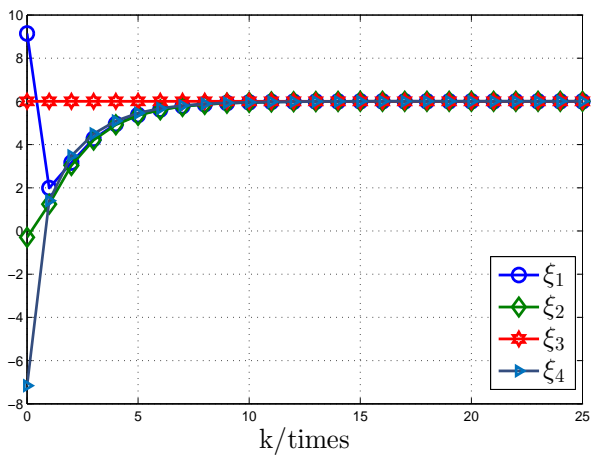}\\
\caption{Opinion evolution of four social actors according to the DeGroot model where the initial opinions are randomly selected from $[-10,10]$.}
\label{20191_fig6}
\end{figure}

\begin{figure}
\centering
\includegraphics[width=3.75in,height=2.5in]{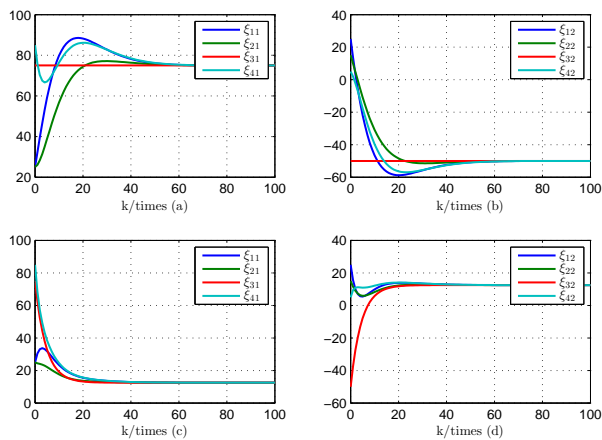}\\
\caption{(a)-(b) Opinion evolution with the issues independence under the cooperative appraisal network; (c)-(d) Opinion evolution with the issues dependence under the cooperative appraisal network.}
\label{20191_fig3}
\end{figure}

\begin{figure}
\centering
\includegraphics[width=3.75in,height=2.5in]{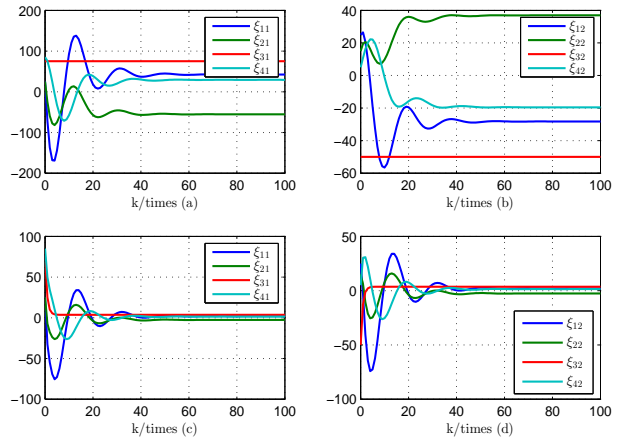}\\
\caption{(a)-(b) Opinion evolution with the issues independence under the antagonistic appraisal network; (c)-(d) Opinion evolution with the issues dependence under the antagonistic appraisal network.}
\label{20191_fig4}
\end{figure}

\begin{figure}
\centering
\includegraphics[width=3.75in,height=2.5in]{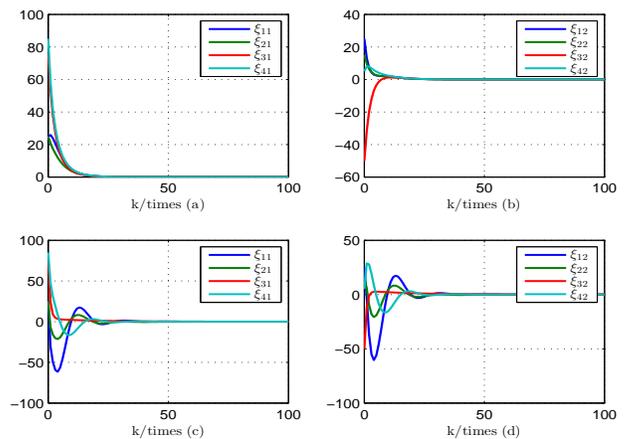}\\
\caption{(a)-(b) Stability of the agents with the cooperative appraisal network depicted in Section ${\rm \ref{20191 sec51}}$; (c)-(d) Stability of the agents with the antagonistic appraisal network depicted in Section ${\rm \ref{20191 sec52}}$.}
\label{20191_fig5}
\end{figure}

\section{Numerical Example}\label{20191 sec5}
Consider a social network with four individuals. The interacting matrix ($P$ corresponds to the DeGroot model) is
\begin{equation}\label{20191eq40}
\begin{aligned}
P=\begin{bmatrix}
0.22 &0.12 &0.36 &0.3\\
0.147 &0.215& 0.344 &0.294\\
0 &0 &1 &0\\
0.09 &0.178 &0.446 &0.286
\end{bmatrix}
\end{aligned}
\end{equation}
It should be pointed out that the elements of matrix $P$ in $(\ref{20191eq40})$ come from real experiment (see \cite{friedkin1999social}). More specific, each entry of $P$ denotes the inter-personal influence by individual $i$ to individual $j$. By Proposition $\ref{20191pro1}$, the Laplacian matrix is
\begin{equation}\label{20191eq44}
\begin{aligned}
\mathcal{L}=\begin{bmatrix}
0.78 &-0.12& -0.36& -0.3\\
-0.147 &0.785& -0.344& -0.294\\
0& 0 &0 &0\\
-0.09 &-0.178 &-0.446& 0.714
\end{bmatrix}
\end{aligned}
\end{equation}
with $\epsilon=1$,  and the associated interacting graph is presented in Fig. $\ref{ch7_interacting_graph}$.  The opinion evolution with respect to four social actors are drawn in Fig. $\ref{20191_fig6}$.
\subsection{Cooperative Appraisal Network}\label{20191 sec51}
In the sequel, we examine the opinion evolution of the interacting agents with the cooperative appraisal network
\begin{equation}\label{20191eq41}
\begin{aligned}
\mathcal {D}_{1}=\begin{bmatrix}
0.2 &0.2 &0.3 &0.3\\
0.1 &0.5 &0 &0.4\\
0.1& 0.4& 0& 0.5\\
 0.4 &0.3 &0.2& 0.1
\end{bmatrix}
\end{aligned}
\end{equation}
The opinion evolution of the individuals for the issue independence ($C_{1}=I$) is plotted in Fig. $\ref{20191_fig3}$(a)-(b) with the susceptibility factor matrix $\Lambda_{1}={\rm diag}(-1, 1, 1 ,-1)$, where the initial opinions are borrowed from \cite[Equation (15)]{parsegov2017novel},
\begin{equation*}\label{20191eq57}
\begin{aligned}
\xi(0)\in \bigg\{\overbrace{25,25}^{\xi_{1}(0)},\overbrace{25,15}^{\xi_{2}(0)},\overbrace{75,-50}^{\xi_{3}(0)},\overbrace{85,5}^{\xi_{4}(0)}\bigg\}
\end{aligned}
\end{equation*}
From Fig. $\ref{20191_fig3}$(a)-(b), the opinions of agents aggregate to the opinion of the leader's, i.e., the $3$rd agent.

In what follows, we discuss the opinion evolution of the interacting agents with the issue interdependence influence. Matrix $C_{1}$ has the form as that in \cite[Section \uppercase\expandafter{\romannumeral7}]{parsegov2017novel},
\begin{equation*}\label{20191eq58}
\begin{aligned}
C_{1}=\begin{bmatrix}
0.9&0.1\\
0.1&0.9
\end{bmatrix}
\end{aligned}
\end{equation*}
With other parameters unchanged, the opinion evolution of agents is depicted in Fig. $\ref{20191_fig3}$(c)-(d). Analogous to the issues independence case, the opinions achieve the consensus. However, some interesting phenomena arise: $\textbf{(1)}$ Unlike the issue free case, the leader's opinion varies over time. Traditionally, the leader's opinion remains the same even if the JF model (including the issue interdependence version \cite{parsegov2017novel}) is applied; $\textbf{(2)}$ By introducing the issue dependence matrix $C_{1}$, the final aggregated opinion may be steered to an opposite direction of the leader's initial opinion. In one word, cooperative appraisal network leads to consensus in opinions, which coincides with the conclusion drawn in Theorem $\ref{20191th1}$.

\subsection{Antagonistic Appraisal Network}\label{20191 sec52}
This subsection focuses on the opinion evolution with antagonistic appraisal network, which is of the form
\begin{equation*}\label{20191eq42}
\begin{aligned}
\mathcal {D}_{2}=\begin{bmatrix}
0.2 &-0.2& -0.3& -0.3\\
0.1 &0.5 &0 &0.4\\
-0.1& 0.4& 0 &0.5\\
0.4& 0.3& -0.2 &0.1
\end{bmatrix}
\end{aligned}
\end{equation*}
Similar to Subsection ${\rm \ref{20191 sec51}}$, we first consider the issue free case. The agents' opinion evolutions are shown in Fig. $\ref{20191_fig4}$(a)-(b) where the parameters are the same except for $\mathcal {D}_{2}$ and $\Lambda_{2}={\rm diag}(-1.5, 2, 1, -0.5)$. Using the method in \cite{parsegov2017novel}, the agents' final opinions share the same direction with the leader's, see \cite[Fig. 5(a)]{parsegov2017novel}. However, the second agent's opinion has an opposite sign with the leader's, even if their initial opinions have the same direction (see Fig. $\ref{20191_fig4}$(a)-(b)).

We proceed with the issue interdependence case with
\begin{equation}\label{20191eq59}
\begin{aligned}
C_{2}=\begin{bmatrix}
0.6& 0.4\\
0.3& 0.7
\end{bmatrix}
\end{aligned}
\end{equation}
Using the same parameters as before, the opinions of the agents are depicted in Fig. $\ref{20191_fig4}$(c)-(d). It is clear that the leader's opinion changes along with the time-evolution, as opposed to the issue free case. In addition, we can see that although some agents have the same direction on the initial opinion of the leader's at the beginning, the agents' final opinions appear an opposite direction with the leader's. Furthermore, the opinions tend to clusters in such a case, which is in accordance with Theorem $\ref{20191th2}$. One can check that $C_{2}$ in $(\ref{20191eq59})$) does not satisfy the conditions in Theorem $\ref{20191th3}$. Indeed, we compute the eigenvalues of matrix $C_{2}$, i.e., $\lambda(C_{2})\in \{1,0.3\}$. They, however, meet the requirement in Theorem $\ref{20191th4}$.

\subsection{Stability of Interacting Agents}\label{20191 sec53}
Now we are dedicated to studying the stability of agents with cooperative and antagonistic appraisal networks. For the cooperative appraisal network, we perform the simulation using the parameters in Subsection ${\rm \ref{20191 sec51}}$ by replacing matrix $C_{1}$ with $C^{\star}_{1}$
\begin{equation*}\label{20191eq58_1}
\begin{aligned}
C^{\star}_{1}=0.85\begin{bmatrix}
0.9&0.1\\
0.1&0.9
\end{bmatrix}
\end{aligned}
\end{equation*}
 It can be verified that the requirement in Theorem $\ref{20191th3}$ is fulfilled. The opinions of the agents are depicted in Fig. $\ref{20191_fig5}$(a)-(b). For the case of antagonistic appraisal network, we use the same parameters as Subsection ${\rm \ref{20191 sec52}}$ by replacing matrix $C_{2}$ with $C^{\star}_{2}$
 \begin{equation*}\label{20191eq59_1}
 \begin{aligned}
 C^{\star}_{2}=0.95\begin{bmatrix}
 0.6& 0.4\\
 0.3& 0.7
 \end{bmatrix}
 \end{aligned}
 \end{equation*}
The opinion evolution of the agents can be found in Fig. $\ref{20191_fig5}$(c)-(d).

\subsection{Further Discussion}\label{20191 sec54}
With the framework of $(\ref{20191eq1})$, we can achieve both consensus and clusters in opinions, and the consensus in opinions is not necessarily restricted into a convex hull as the classical DeGroot model (see Fig. $\ref{20191_fig2}$(a)). Moreover, formulation $(\ref{20191eq1})$ extends the protocol of \cite{zhang2017quasi} in several aspects: ({\rm \lowercase \expandafter {\romannumeral 1}}) it could be utilized to describe more general behaviors in social networks; ({\rm \lowercase \expandafter {\romannumeral 2}}) it further manifests the importance on the weighted gain matrix $\Lambda$. In other words, system $(\ref{20191eq2})$ may diverge without the help of $\Lambda$.

In \cite[Equation (15)]{parsegov2017novel}, Parsegov \emph{et al}. endowed both the initial opinions and the final opinions of the social actors with some specific meanings: the positive (resp. the negative) opinions correspond to the vegetarian (resp. the all-meat) diets by introducing an interdependent issue matrix. According to \cite[Examples $3$ and $4$]{parsegov2017novel}, all social actors are the vegetarian which is coincide with their initial opinions, $\xi^{1}(0)=(25,25,75,85)$ (cf. \cite[Equation (15)]{parsegov2017novel}), i.e., the initial opinions on the first issue. However, we can see from Fig. $\ref{20191_fig3}$, some agents become the all-meat diets eventually, even if they are the vegetarian at the beginning. This also indicates that some agents have the opposite attitudes compared with the leader's ever though they all have the same direction of attitudes at first. Additionally, from Figs. $\ref{20191_fig3}$(c)-(d) and $\ref{20191_fig4}$(c)-(d), the leader's final opinion may be affected by the evolutions of other agents, which is new from the perspective of the classical DeGroot model. In a nutshell, the proposed setup in this paper brings some interesting phenomena comparing with the existing literature.

\section{Conclusion}\label{20191 sec6}
This paper has studied the opinion dynamics in social networks by introducing an appraisal network to quantify the cooperative or antagonistic information. We have shown that the cooperative appraisal network achieves the consensus in opinions while the antagonistic appraisal network leads to the opinion clusters. The tool of random convex optimization is used to estimate the appraisal network with a confident level of robustness, along with the lower bound on the amounts of sampled observations. Moreover, the proposed setup has been extended to the case of multiple issues interdependence. Some discussions have also been given to compare with the existing literature.

\appendix[Proof of Theorem $\ref{20191th6}$]
To prove Theorem $\ref{20191th6}$, the Hermite-Biehler Theorem (cf. \cite{long1992robust}) and the Bilinear Transformation Theorem (cf. \cite{ogata1995discrete}) are needed. To begin with, we first introduce a lemma.
\begin{lem}(\hspace{-0.001cm}\cite{ogata1995discrete})\label{20191le4}
Given two polynomials $\mathbb{S}(z)$ of degree $d$ and $\mathbb{Q}(z)$ with
\begin{equation*}\label{20191eq68}
\begin{aligned}
\mathbb{Q}(z)=~(z-1)^{d}\mathbb{S}\bigg(\frac{z+1}{z-1}\bigg)
\end{aligned}
\end{equation*}
Then the Schur stability of $\mathbb{S}(z)$ implies the Hurwitz stability on $\mathbb{Q}(z)$, and vice versa.
\end{lem}

For complex polynomial $\mathbb{Q}(z)$, replacing $z$ with $\mathbbm{i} w$ yields
\begin{equation*}\label{20191eq69}
\begin{aligned}
\mathbb{Q}(\mathbbm{i} w)=~S(w)+\mathbbm{i} Q(w)
\end{aligned}
\end{equation*}
where both $S(w)$ and $Q(w)$ are real polynomials. A relationship among the roots of $S(w)$ and $Q(w)$ is depicted as follows.

\begin{de}(\hspace{-0.001cm}\cite{long1992robust})\label{20191de3}
For any real polynomials $S(w)$ and $Q(w)$, they are interlaced if
\begin{itemize}
\item[({\rm \lowercase \expandafter {\romannumeral 1}})] The roots of $S(w)$ (denoted by $\{S_{1},...,s_{\ell_{S}}\}$) and $Q(w)$ (denoted by $\{Q_{1},...,Q_{\ell_{Q}}\}$) satisfy
    \begin{equation*}\label{20191eq70}
\left\{\begin{aligned}
&S_{1}<S_{2}<\cdots<S_{\ell_{S}}\\
&Q_{1}<Q_{2}<\cdots<Q_{\ell_{Q}}
\end{aligned}\right.
\end{equation*}
\item[({\rm \lowercase \expandafter {\romannumeral 2}})]
$|\ell_{S}-\ell_{Q}|\leq1$, and one of the following facts holds
\begin{equation*}\label{20191eq71}
\left\{\begin{aligned}
&S_{1}<Q_{1}<S_{2}<Q_{2}\cdots<Q_{\ell_{Q}}<S_{\ell_{S}}, ~ \ell_{S}=\ell_{Q}+1\\
&Q_{1}<S_{1}<Q_{2}<S_{2}\cdots<S_{\ell_{S}}<Q_{\ell_{Q}}, ~ \ell_{Q}=\ell_{S}+1\\
&Q_{1}<S_{1}<Q_{2}<S_{2}\cdots<Q_{\ell_{Q}}<S_{\ell_{S}}, ~ \ell_{Q}=\ell_{S}\\
&S_{1}<Q_{1}<S_{2}<Q_{2}\cdots<S_{\ell_{S}}<Q_{\ell_{Q}}, ~ \ell_{S}=\ell_{Q}\\
\end{aligned}\right.
\end{equation*}
\end{itemize}
\end{de}

The next lemma develops a judgement on the Hurwitz stability associated with $\mathbb{Q}(z)$ that is closely linked to the roots of polynomials $S(w)$ and $Q(w)$ and the interlaced property defined in Definition $\ref{20191de3}$.

\begin{lem}(\hspace{-0.001cm}\cite{long1992robust})\label{20191le5}
Polynomial $\mathbb{Q}(z)$ is Hurwitz stable if and only if
\begin{itemize}
\item[({\rm \lowercase \expandafter {\romannumeral 1}})] Polynomials $S(w)$ and $Q(w)$ are interlaced;
\item[({\rm \lowercase \expandafter {\romannumeral 2}})] Polynomials $S(w)$, $\frac{\partial Q(w)}{\partial w}$, $\frac{\partial S(w)}{\partial w}$ and $Q(w)$ at the origin fulfill
\begin{equation*}\label{20191eq72}
\begin{aligned}
S(0)\frac{\partial Q(0)}{\partial w}-\frac{\partial S(0)}{\partial w}Q(0)>0
\end{aligned}
\end{equation*}
\end{itemize}
\end{lem}

With the above preparations, we are about to give the proof of Theorem $\ref{20191th6}$.

\begin{IEEEproof}[Proof of Theorem $\ref{20191th6}$]
It is easy to see that the updating dynamics of the individuals is determined by $(\ref{20191eq64})$.
Accordingly, the characteristic equation of $(\ref{20191eq64})$ can be written as
\begin{equation*}\label{20191eq74}
\begin{aligned}
&{\rm det}\bigg(z I-(I-\varrho\mathcal {L}+\varrho^{2}\mathcal {L}^{2})\bigg)\\
=&~(z-1)\prod^{N}_{i=2}(z-1+\varrho \lambda_{i}-
\varrho^{2} \lambda^{2}_{i})
\end{aligned}
\end{equation*}
Before proceeding further, it is indispensable to consider two cases where ${\rm Im} (\lambda_{i})$ is equal to zero or not. And we start with the scenario that ${\rm Im} (\lambda_{i})=0$.

As argued before, consensus in opinions indicates that $|z|<1$ if $z\neq1$. In such a setting, it is enough to demonstrate that
\begin{subequations}\label{20191eq79}
\begin{empheq}[left=\empheqlbrace]{align}
&\varrho^{2} \lambda^{2}_{i}-\varrho \lambda_{i}+1<1 \label{20191eq79a}\\
&\varrho^{2} \lambda^{2}_{i}-\varrho \lambda_{i}+1>-1  \label{20191eq79b}
\end{empheq}
\end{subequations}
with the constraints $\varrho>0$ and $\lambda_{i}\neq 0$. By computation, one derives that the feasible region on $\varrho$ for inequality $(\ref{20191eq79a})$ is $(0, \frac{1}{\lambda_{i}})$, while the feasible region on $\varrho$ associated with inequality $(\ref{20191eq79b})$ is $[0, \infty)$. Thus, the feasible region for $\varrho$ is
\begin{equation*}\label{20191eq80}
\begin{aligned}
\varrho \in \bigg(0,\frac{1}{\lambda_{i}}\bigg)
\end{aligned}
\end{equation*}

In the sequel, we are dedicated to the case of ${\rm Im} (\lambda_{i})\neq0$. Let $\mathbb{S}_{i}(z)$ be of the form
\begin{equation}\label{20191eq75}
\begin{aligned}
\mathbb{S}_{i}(z)=~z-1+\varrho \lambda_{i}-
\varrho^{2} \lambda^{2}_{i}, ~i=2,...,N
\end{aligned}
\end{equation}
where $\lambda_{i}$ stands for the $i$th eigenvalue of $\mathcal {L}$. Applying the bilinear transformation
\begin{equation}\label{20191eq76}
\begin{aligned}
\mathbb{Q}_{i}(z)=~(z-1)\mathbb{S}_{i}\bigg(\frac{z+1}{z-1}\bigg)
\end{aligned}
\end{equation}
One gets
\begin{equation*}\label{20191eq77}
\begin{aligned}
\mathbb{Q}_{i}(z)=&~2+\varrho\lambda_{i}(1-\varrho\lambda_{i})z+\varrho\lambda_{i}(\varrho\lambda_{i}-1)\\
=&~2+\varrho|\lambda_{i}|\bigg(\cos(\arg(\lambda_{i}))+\mathbbm{i}\sin(\arg(\lambda_{i}))\bigg)z\\
&-\varrho|\lambda_{i}|\bigg(\cos(\arg(\lambda_{i}))+\mathbbm{i}\sin(\arg(\lambda_{i}))\bigg)\\
&-\varrho^{2} |\lambda_{i}|^{2}\bigg(\cos^{2}(\arg(\lambda_{i}))-\sin^{2}(\arg(\lambda_{i}))\\
&+2\mathbbm{i}\sin(\arg(\lambda_{i}))\cos(\arg(\lambda_{i}))\bigg)z\\
&-\varrho^{2} |\lambda_{i}|^{2}\bigg(\cos^{2}(\arg(\lambda_{i}))-\sin^{2}(\arg(\lambda_{i}))\\
&+2\mathbbm{i}\sin(\arg(\lambda_{i}))\cos(\arg(\lambda_{i}))\bigg)\\
\end{aligned}
\end{equation*}
Substituting $z=\mathbbm{i}w$ into $\mathbb{Q}_{i}(z)$ results in
\begin{equation}\label{20191eq78}
\begin{aligned}
\mathbb{Q}_{i}(\mathbbm{i}w)=&~~2+\varrho|\lambda_{i}|\bigg(\mathbbm{i}\cos(\arg(\lambda_{i}))-\sin(\arg(\lambda_{i}))\bigg)w\\
&-\varrho|\lambda_{i}|\bigg(\cos(\arg(\lambda_{i}))+\mathbbm{i}\sin(\arg(\lambda_{i}))\bigg)\\
&-\varrho^{2} |\lambda_{i}|^{2}\bigg(\mathbbm{i}\cos^{2}(\arg(\lambda_{i}))-\mathbbm{i}\sin^{2}(\arg(\lambda_{i}))\\
&-2\sin(\arg(\lambda_{i}))\cos(\arg(\lambda_{i}))\bigg)w\\
&-\varrho^{2} |\lambda_{i}|^{2}\bigg(\cos^{2}(\arg(\lambda_{i}))-\sin^{2}(\arg(\lambda_{i}))\\
&+2\mathbbm{i}\sin(\arg(\lambda_{i}))\cos(\arg(\lambda_{i}))\bigg)\\
\triangleq&~{\rm Re}(\mathbb{Q}_{i}(w))+\mathbbm{i}{\rm Im}(\mathbb{Q}_{i}(w))
\end{aligned}
\end{equation}
With the constraint ${\rm Re}(\mathbb{Q}_{i}(w))=0$, it yields
\begin{equation*}\label{20191eq81}
\begin{aligned}
&w_{1}\\
=&~\frac{2+\varrho^{2} |\lambda_{i}|^{2}(2\cos^{2}(\arg(\lambda_{i}))-1)-\varrho|\lambda_{i}|\cos(\arg(\lambda_{i}))}
{\varrho|\lambda_{i}|\sin(\arg(\lambda_{i}))-2\varrho^{2} |\lambda_{i}|^{2}\sin(\arg(\lambda_{i}))\cos(\arg(\lambda_{i}))}\\
\triangleq&~\frac{2+\mathbbm{x}}{\mathbbm{y}}
\end{aligned}
\end{equation*}
Analogously, the requirement of ${\rm Im}(\mathbb{Q}_{i}(w))=0$ immediately leads to
\begin{equation*}\label{20191eq82}
\begin{aligned}
&w_{2}\\
=&~\frac{2\varrho^{2} |\lambda_{i}|^{2}\sin(\arg(\lambda_{i}))\cos(\arg(\lambda_{i}))-\varrho|\lambda_{i}|\sin(\arg(\lambda_{i}))}
{\varrho^{2} |\lambda_{i}|^{2}(2\cos^{2}(\arg(\lambda_{i}))-1)-\varrho|\lambda_{i}|\cos(\arg(\lambda_{i}))}
\end{aligned}
\end{equation*}
In accordance with $w_{1}$, $w_{2}$ hence is of the form
\begin{equation*}\label{20191eq100}
\begin{aligned}
w_{2}=~\frac{-\mathbbm{y}}{\mathbbm{x}}
\end{aligned}
\end{equation*}
We now determine the condition guaranteeing
\begin{equation*}\label{20191eq83}
\begin{aligned}
w_{1}\frac{1}{w_{2}}\neq 1
\end{aligned}
\end{equation*}
For ease of discussion, we resort to the opposite statement, i.e.,
\begin{equation}\label{20191eq84}
\begin{aligned}
1=&~w_{1}\frac{1}{w_{2}}\\
=&~\frac{2\mathbbm{x}+\mathbbm{x}^{2}}{-\mathbbm{y}^{2}}
\end{aligned}
\end{equation}
We can see that $(\ref{20191eq84})$ is equivalent to
\begin{equation*}
\begin{aligned}
1=~(\mathbbm{x}+1)^{2}+\mathbbm{y}^{2}
\end{aligned}
\end{equation*}

By the polar coordination, for $\theta\in[0,2\pi) $, it attains
\begin{subequations}\label{20191eq85}
\begin{empheq}[left=\empheqlbrace]{align}
\mathbbm{x}=&~\cos(\theta)-1 \label{20191eq85a}\\
\mathbbm{y}=&~\sin(\theta) \label{20191eq85b}
\end{empheq}
\end{subequations}
For $(\ref{20191eq85a})$, it is more prone to access that
\begin{equation*}\label{20191eq86}
\begin{aligned}
\mathbbm{x}-\cos(\theta)+1=&~\varrho^{2}|\lambda_{i}|^{2}\cos(2\arg(\lambda_{i}))\\
&-\varrho|\lambda_{i}|\cos(\arg(\lambda_{i}))-\cos(\theta)+1\\
\triangleq&~f_{\theta}(\varrho)
\end{aligned}
\end{equation*}
For $\forall~\theta\in[0,2\pi)$, denote
\begin{equation*}
\begin{aligned}
\Delta_{\mathbbm{x}}\triangleq &~|\lambda_{i}|^{2}+4|\lambda_{i}|^{2}\cos(2\arg(\lambda_{i}))(\cos(\theta)-1)\\
=&~|\lambda_{i}|^{2}\bigg(\cos^{2}(\arg(\lambda_{i}))(8\cos(\theta)-7)+4(1-\cos(\theta))\bigg)
\end{aligned}
\end{equation*}
In the sequel, two cases involving the real roots of $f_{\theta}(\varrho)=0$ are formulated.

Case 1) $\theta\in[-\arccos(\frac{7}{8}),\arccos(\frac{7}{8})]$, $\Delta_{\mathbbm{x}}\geq0$ follows directly.

Case 2) $\theta\in(-\frac{\pi}{2},-\arccos(\frac{7}{8}))\bigcup(\arccos(\frac{7}{8}),\frac{\pi}{2})$. In such a circumstance, one has
$\Delta_{\mathbbm{x}}\geq0$ if and only if $\arg(\lambda_{i})$ is preserved in the set
\begin{equation*}\label{20191eq92}
\begin{aligned}
&\bigg[ \arccos( \sqrt{\frac{4(1-\cos(\theta))}{7-8\cos(
\theta)}} ,   \frac{\pi}{2})\bigg)\\
\bigcup&
 \bigg(-\frac{\pi}{2}, -\arccos( \sqrt{\frac{4(1-\cos(\theta))}{7-8\cos(
\theta)}}  )  \bigg]
\end{aligned}
\end{equation*}
The two real roots of $f_{\theta}(\varrho)=0$ are given by
\begin{equation*}\label{20191eq87}
\left\{\begin{aligned}
\varrho_{i,1}=&~\frac{|\lambda_{i}|\cos(\arg(\lambda_{i}))+\sqrt{\Delta_{\mathbbm{x}}}}{2|\lambda_{i}|^{2} \cos(2\arg(\lambda_{i}))    }\\
\varrho_{i,2}=&~\frac{|\lambda_{i}|\cos(\arg(\lambda_{i}))-\sqrt{\Delta_{\mathbbm{x}}}}{2|\lambda_{i}|^{2} \cos(2\arg(\lambda_{i}))    }\\
\end{aligned}\right.
\end{equation*}
Moreover, we calculate
\begin{equation*}\label{20191eq88}
\begin{aligned}
|\lambda_{i}|^{2}\cos^{2}(\arg(\lambda_{i}))-\Delta_{\mathbbm{x}}=4|\lambda_{i}|^{2}(1-\cos(\theta))\cos(2 \arg(\lambda_{i}))
\end{aligned}
\end{equation*}
Therefore, the sign of the calculation value is entirely contingent on $\cos(2 \arg(\lambda_{i}))$, which in turn indicates that both $\varrho_{1}$ and $\varrho_{2}$ are positive.

For $(\ref{20191eq85b})$, we find
\begin{equation*}\label{20191eq89}
\begin{aligned}
\sin(\theta)-\mathbbm{y}=&~2\varrho^{2}|\lambda_{i}|^{2}\sin(2\arg(\lambda_{i}))
-\varrho|\lambda_{i}|\sin(\arg(\lambda_{i})\\
&+\sin(\theta)\\
\triangleq&~g_{\theta}(\varrho),~\theta\in[0,2\pi)
\end{aligned}
\end{equation*}
For $\forall~\theta\in[0,2\pi)$, denote
\begin{equation*}\label{20191eq90}
\begin{aligned}
\Delta_{\mathbbm{y}}\triangleq &~|\lambda_{i}|^{2}\bigg
(\sin^{2}(\arg(\lambda_{i}))-4\sin(2\arg(\lambda_{i}))\sin(\theta)\bigg)
\end{aligned}
\end{equation*}
Evidently, for $\theta\in[0,2\pi)$, $\Delta_{\mathbbm{y}}\geq0$ implies
\begin{equation}\label{20191eq91}
\begin{aligned}
\sin^{2}(\arg(\lambda_{i}))\geq 8\sin(\arg(\lambda_{i}))\cos(\arg(\lambda_{i}))\sin(\theta)
\end{aligned}
\end{equation}
In a similar manner, two scenarios should be argued.

Case {\rm \lowercase \expandafter {\romannumeral 1}}) $\arg(\lambda_{i})\in(0,\frac{\pi}{2})$. In this case, $(\ref{20191eq91})$ is desirable provided that
\begin{equation*}\label{20191eq101}
\begin{aligned}
\arg(\lambda_{i})\in \bigg[\max\{0,\arctan(8\sin(\theta))\},\frac{\pi}{2}\bigg) \backslash \{ 0\},~\theta\in[0,2\pi)
\end{aligned}
\end{equation*}

Case {\rm \lowercase \expandafter {\romannumeral 2}}) $\arg(\lambda_{i})\in(-\frac{\pi}{2},0)$. It is trivial that
\begin{equation*}\label{20191eq102}
\begin{aligned}
0\leq \Delta_{\mathbbm{y}}, ~\forall~\theta \in [0,\pi]
\end{aligned}
\end{equation*}
And for $\theta \in [-\pi,0]$, $(\ref{20191eq91})$ could be further expressed by
\begin{equation*}\label{20191eq103}
\begin{aligned}
\sin^{2}(\arg(\lambda_{i}))\geq 8|\sin(\arg(\lambda_{i}))|\cos(\arg(\lambda_{i}))|\sin(\theta)|
\end{aligned}
\end{equation*}
which is true as long as
\begin{equation*}\label{20191eq104}
\begin{aligned}
\arg(\lambda_{i})\in \bigg[-\frac{\pi}{2}, \min\{0,-\arctan(8\sin(\theta))\}\bigg) \backslash \{ 0\}
\end{aligned}
\end{equation*}
With the foregoing arguments, $g_{\theta}(\varrho)=0$ has two real roots, which can be described as
\begin{equation*}\label{20191eq105}
\left\{\begin{aligned}
\varrho_{i,3}=&~\frac{|\lambda_{i}|\sin(\arg(\lambda_{i}))+\sqrt{\Delta_{\mathbbm{y}}}}{2|\lambda_{i}|^{2} \sin(2\arg(\lambda_{i}))    }\\
\varrho_{i,4}=&~\frac{|\lambda_{i}|\sin(\arg(\lambda_{i}))-\sqrt{\Delta_{\mathbbm{y}}}}{2|\lambda_{i}|^{2} \sin(2\arg(\lambda_{i}))    }\\\end{aligned}\right.
\end{equation*}
Therefore, $w_{1}$ and $w_{2}$ are interlaced (cf. Definition $\ref{20191de3}$) with the constraints on $\varrho$, i.e.,
\begin{equation*}\label{20191eq106}
\begin{aligned}
\varrho\not\in\{\varrho_{i,1},\varrho_{i,2}\}\bigcup \{\varrho_{i,3},\varrho_{i,4}\}
\end{aligned}
\end{equation*}

By virtue of the specifications on ${\rm Re}(\mathbb{Q}_{i}(w))$ and ${\rm Im}(\mathbb{Q}_{i}(w))$ (cf. $(\ref{20191eq78})$), one gets
\begin{equation*}\label{20191eq93}
\left\{\begin{aligned}
\frac{\partial{\rm Re}(\mathbb{Q}_{i}(0))}{\partial w}=&~\varrho^{2} |\lambda_{i}|^{2}\sin(2\arg(\lambda_{i}))-\varrho |\lambda_{i}|\sin(\arg(\lambda_{i}))\\
{\rm Re}(\mathbb{Q}_{i}(0))=&~2+\varrho^{2} |\lambda_{i}|^{2}(2\cos^{2}(\arg(\lambda_{i}))-1)\\
&-\varrho|\lambda_{i}|\cos(\arg(\lambda_{i}))\\
\frac{\partial{\rm Im}(\mathbb{Q}_{i}(0))}{\partial w}=&~\varrho^{2} |\lambda_{i}|^{2}(1-2\cos^{2}(\arg(\lambda_{i})))\\
&+\varrho|\lambda_{i}|\cos(\arg(\lambda_{i}))\\
{\rm Im}(\mathbb{Q}_{i}(0))=&~\varrho^{2} |\lambda_{i}|^{2}\sin(2\arg(\lambda_{i}))-\varrho |\lambda_{i}|\sin(\arg(\lambda_{i}))
\end{aligned}\right.
\end{equation*}
We perform the calculation
\begin{equation*}\label{20191eq94}
\begin{aligned}
&{\rm Re}(\mathbb{Q}_{i}(0))\frac{\partial{\rm Im}(\mathbb{Q}_{i}(0))}{\partial w}-\frac{\partial{\rm Re}(\mathbb{Q}_{i}(0))}{\partial w}{\rm Re}(\mathbb{Q}_{i}(0))\\
=&~\varrho |\lambda_{i}|\bigg( -\varrho^{3} |\lambda_{i}|^{3}+\varrho^{2} |\lambda_{i}|^{2} \cos^{2}(\arg(\lambda_{i}))\\
&-2\varrho |\lambda_{i}|\sin(2\arg(\lambda_{i})) -\varrho |\lambda_{i}|+2\cos(\arg(\lambda_{i})) \bigg)\\
\triangleq&~\varrho |\lambda_{i}| f_{i}(\varrho,\lambda_{i},\arg(\lambda_{i})),~i=2,...,N
\end{aligned}
\end{equation*}
Obviously, ${\rm Re}(\mathbb{Q}_{i}(0))\frac{\partial{\rm Im}(\mathbb{Q}_{i}(0))}{\partial w}-\frac{\partial{\rm Re}(\mathbb{Q}_{i}(0))}{\partial w}{\rm Re}(\mathbb{Q}_{i}(0))$ is positive if and only if $f_{i}(\varrho,\lambda_{i},\arg(\lambda_{i}))>0$.

Consequently, by Lemma $\ref{20191le5}$, we can see that $\mathbb{Q}_{i}(z)$ in $(\ref{20191eq76})$ is Hurwitz stable. With the aid of Lemma $\ref{20191le4}$, one could state that $\mathbb{S}_{i}(z)$ in $(\ref{20191eq75})$ is Schur stable. This ends the proof.
\end{IEEEproof}

\end{document}